\documentclass{emulateapj}
\usepackage{natbib,graphicx,url}
\shorttitle{Nanolensing with Subsolar Mass Halos}
\shortauthors{Chen \& Koushiappas}

\begin{document}

\title{Gravitational Nanolensing from subsolar mass dark matter halos}
\author{Jacqueline Chen\altaffilmark{1,2} and Savvas M. Koushiappas\altaffilmark{1,3}}
\altaffiltext{1}{Department of Physics, Brown University, 182 Hope St., Box 1843, Providence, RI 02906} 
\altaffiltext{2}{{\tt Jacqueline\_Chen@brown.edu}} 
\altaffiltext{3}{{\tt koushiappas@brown.edu}} 

\begin{abstract}
We investigate the feasibility of extracting the gravitational nanolensing signal due to the presence of subsolar mass halos within galaxy-sized dark matter halos. We show that subsolar mass halos in a lensing galaxy can cause strong nanolensing events with shorter durations and smaller amplitudes than microlensing events caused by stars. We develop techniques that can be used in future surveys such as Pan-STARRS, LSST and OMEGA to search for the nanolensing signal from subsolar mass halos. 
\end{abstract}


\section{Introduction}

In cold dark matter (CDM) cosmologies, dark matter halos form in a hierarchical manner. Small dark matter halos form first, and their subsequent merger results in the formation of larger systems. Dark matter halos contain a remnant of this process in the form of substructure, the self-bound smaller systems that survive the hierarchical assembly of their host halo. 

The scale of the first objects that form in the Universe is set by the nature of the dark matter particle. An observation of a cutoff scale in the dark matter power spectrum (either in the present-day Universe or in earlier epochs) would provide an insight into the particle physics properties of the dark matter, as well as information about the dynamics and evolution of non-linear systems at the dawn of structure formation. 

On scales greater than $\sim 10^7 M_\odot$, luminous galaxies form in the center of the dark matter potential well (e.g., dwarf galaxies, galaxies, and clusters of galaxies).  On scales less than $\sim 10^7 M_\odot$ the collisional nature of baryons prevents the formation of visible objects, making the use of luminous matter as a tracer of the dark matter distribution ineffective \citep{2008Natur.454.1096S,2010AdAst2010E...8K,2010arXiv1003.4268S}. Thus, the existence of self-bound dark matter systems with masses less than approximately $10^7 M_\odot$ is very difficult to establish. This presents a very serious problem in attempts to understand the small-scale structure of dark matter halos. 

Weakly Interacting Massive Particles (WIMPs) are well established theoretically and offer a natural dark matter relic density contribution \citep{Jungman:1995df,BHS05}. They arise in theories which introduce physics beyond the Standard Model and have masses of order of tens of GeV and an interaction strength with Standard Model particles of the order of the weak interaction. The free streaming of WIMPs after kinetic decoupling in the early Universe (momentum-changing interactions) introduces a cutoff in the matter power spectrum typically at a sub-solar mass scale below $\approx 10^{-4} M_{\sun}$ \citep{Schmid:1998mx,Green:2003un,Green:2005fa}. This cutoff scale has been theoretically explored in the particular case of supersymmetric WIMPs \citep{HSS01, CKZ02,Profumo:2006bv,LZ05,2009JCAP...06..014M}. While initially the first dark matter halos had masses near the cutoff scale, it is unclear whether these objects survived the hierarchical growth of their host halo \citep{DMS05,Berezinsky:2003vn,Green:2006hh,Goerdt:2006hp,ishiyama_etal10}. Numerical simulations, as well as sophisticated analytical arguments, cannot at present address this issue conclusively, and therefore any potential observational signature of the presence of subsolar mass halos is valuable. 

The possibility of the detection of subsolar mass dark matter halos by indirect and direct detection experiments has been studied extensively in literature. Indirect detection experiments search for the products of dark matter annihilation. As this process is proportional to the integral of the square of the dark matter density distribution over volume, high density regions enhance the annihilation rate relative to low density regions. Subsolar mass dark matter halos have very high densities as they are formed at extremely high redshifts (their densities reflect the mean dark matter density of the Universe at the redshift of formation). The probability of detecting the indirect detection effects of subsolar mass subhalos present in present day dark matter halos has been studied in the context of individual detection \citep{diemand_etal05b,Pieri:2005pg,Koushiappas:2006qq}, as well as their contribution to the diffuse gamma-ray background \citep{2008PhRvD..78j1301A,2010PhRvD..81d3532K,Pieri:2007ir,2009JCAP...07..007L,2010arXiv1006.2399B}. In addition, a subject that received considerable attention is the role of subsolar mass substructure in setting the amplitude of the annihilation flux in halos, the so-called ``boost" factor \citep{Bergstrom:1998jj,Bergstrom:1998zs,CalcaneoRoldan:2000yt,Berezinsky:2003vn,Stoehr:2003hf,Baltz:2006sv,Diemand:2006ik,Berezinsky:2006qm,2007PhRvD..75h3526S,Kuhlen:2008aw,Kuhlen2009kx,Kamionkowski:2010mi}. 

On the other hand, direct detection experiments are minimally sensitive to the presence of subsolar mass halos in the Milky Way.  Inspired by analytical arguments \citep{2008PhRvD..77j3509K}, numerical simulations show that the direct detection experiments are not sensitive to subsolar mass subhalos for two reasons \citep{2009MNRAS.395..797V}.  First, the volume occupied by subsolar mass substructure at the present epoch is extremely small (thus the probability of an interaction with a subsolar mass halo during the course of a direct detection experiment is negligible). Second, the smooth dark matter distribution does not retain any information about the presence of subsolar mass halos (even as tidal streams) due to the very efficient mixing of tidally stripped material since the dynamical timescale at the solar radius is a very small fraction of the age of the Milky Way \citep{2010arXiv1002.3162V}.  

Finally, energetic neutrinos from the Sun and the Earth could potentially hold a signature of their interactions with subsolar mass halos along the Sun's Galactic orbit:   the signal samples the past history of the local dark matter density \citep{2009PhRvL.103l1301K}. A past interaction with a subsolar mass halo could give rise to an elevated signal of energetic neutrinos, though such an effect is sensitive to the equilibration timescale of the dark matter particle, the survival of subsolar mass halos in the Milky Way, and their internal structure \citep{2009PhRvL.103l1301K,2010arXiv1006.3268S}. 

Gravitational lensing offers a new possibility for the detection of subsolar mass dark matter halos.  Light from distant objects is deflected by matter along the line of sight.  In some cases, such as a sufficiently large intervening galaxy, the observer will see multiple images of the same source;  this is referred to as `strong' gravitational lensing.  Strong lens systems with quasar sources and galaxy-sized lenses have been shown to be sensitive to stars \citep{chang_refsdal79,1986ApJ...301..503P,1986A&A...166...36K,1987A&A...171...49S,1989AJ.....98.1989I,wambsganss_etal90,wambsganss_paczynski91}, as well as satellite galaxy-sized dark matter subhalos in the lensing galaxy \citep{mao_schneider98,metcalf_madau01,metcalf_zhao02,dalal_kochanek02,chiba02,kochanek_dalal04}.  The sensitivity of lensing systems to the larger dark matter subhalos is a well-known probe of the nature of the dark matter particle \citep[see, e.g.,][]{moustakas_etal09}. These systems could also be sensitive to much smaller dark matter substructure \citep{lewis_gilmerino06}.  If so, strong gravitational lensing might provide the best estimate for detecting subsolar mass DM halos since the lensing signal is sensitive only to mass and does not depend upon the nature of the dark matter particle.  

In this paper, we focus on this new possibility of using gravitational lensing as a probe of the existence of subsolar mass subhalos in dark matter halos.  In Sec.~\ref{sec:simple} we present an order of magnitude estimate of the effect based on a simple toy model. As the main source of confusion in such an investigation is the presence of stars in the lens, we construct a mock lens system  and investigate the lensing effects of stars and dark matter halos in Sec.~\ref{sec:nano}.   We then create mock observations in Sec.~\ref{sec:mock} and search for the signatures of subsolar mass dark matter halos in Sec.~\ref{sec:analysis}.  In Sec.~\ref{sec:discussion} we discuss challenges to measuring subsolar mass dark matter halos in real observations, and we discuss the prospects for detection in current and future data sets in Sec.~\ref{sec:conclusions}.

\section{Simple Estimates}
\label{sec:simple}

In strong gravitational lens systems, a distant source is multiply-imaged by an intervening object.  In cases for which the source is a quasar and the lens is a galaxy, small sub-galactic objects can cause large perturbations to the magnification of individual images.  Investigating such effects is computationally intensive, so here we discuss in a simplified way the prospects for finding such large magnification perturbations due to subsolar mass dark matter halos.

A simple proxy for large magnification perturbations is the splitting of an image into subimages.  The total magnification of the subimages is at least as big as the magnification in the absence of image splitting.  In addition, multiple subimages imply the existence of solutions of the lens equation for which the magnification of resulting subimages can be arbitrarily high -- i.e., critical points.   

When the perturbers in the lens are stars, multiple subimages are always present if the star and the source are sufficiently aligned because stars are point masses.  Stars cause image splittings at roughly the micro-arcsecond ($\mu$as) scale, giving the effect the name `microlensing.'\footnote{For a comprehensive discussion of microlensing, see \citet{wambsganss06}.}   Subsolar mass dark matter halos, however, are not point masses.  A sufficient condition for multiple subimages is that the surface mass density is greater than the critical density of lensing,
\begin{equation}
\kappa \equiv \kappa_{\rm total} = \Sigma / \Sigma_{\rm crit} \geq 1.
\end{equation} 
Here we specify  $\kappa_{\rm total}$ to emphasize the fact that $\kappa$ can have contributions from many sources, such as subsolar mass halos or a smooth dark matter component.

Let us consider some toy parameters for a subsolar mass halo.  At the redshift of collapse, a $\sim$1 solar mass ($M_{\sun}$) halo has a virial radius of $\sim$1 parsec (pc) \citep{koushiappas09}.  At formation, studies suggest that the halo is described by a \citet[NFW]{navarro_etal97} profile, $\rho(r) \propto 1/[x(1+x)^2]$ where $x=r/r_s$ and $r$ is the three-dimensional radius and $r_s$ is the scale radius \citep{diemand_etal05b,diemand_etal06}.  \citet{ishiyama_etal10} find $\rho \propto r^{-1.5}$ using higher resolution simulations, but ignoring the larger scale fluctuations that could allow halos to form at earlier redshift and with higher central densities.  Let us approximate the projected mass profile of the halo as a power law, $\Sigma(R) \propto R^{ - \alpha}$, where $R$ is the projected radius and $\alpha > 0$.  We consider a lensing galaxy at $z=0.3$ and source quasar at $z=2$.  

At the position of the images in strong lens systems, $\kappa_{\rm total} \sim$ 0.5. If all the mass is in solar mass halos, then $\approx$ 3000 halos will overlap at any given point.   Now let us separate the contribution of one halo from all the other halos so that $\kappa_{\rm total} = \kappa_{\rm 1 halo} + \kappa_{\rm back}$.  Given the high average number density of halos, the average contribution of all but one halo at a point is $\kappa_{\rm back} \sim0.5$.  If we then consider the surface density contribution of a single halo, there is an area very close to the center of this halo  where the surface density is $\kappa_{\rm 1 halo} \geq 0.5$.  This occurs at a radius of 0.002, 0.0002, and $7\times 10^{-8}$ pc for a halo profile slope of $\alpha = 1.5$, $1$, and $0.5$ respectively.  An estimate of the probability of a source point falling into the area of $\kappa_{\rm total} \geq 1$ is 0.03, $3 \times 10^{-4}$, and $4 \times 10^{-11}$ for a halo profile slope of $\alpha = 1.5$, $1$, and $0.5$ respectively.  

The area of $\kappa_{\rm 1 halo} \geq 0.5$ also corresponds roughly to the size of the deflection caused by the subsolar mass dark matter halos and is $\sim0.1$ $\mu$as for $\alpha=1.5$.  The mass function of subsolar masses extends significantly below a solar mass, and the typical deflections for smaller halos will be correspondingly smaller and encompass the nano-arcsecond (nas) scale.  As the lensing effects of stars are generically referred to as `microlensing,' we refer to the effects of subsolar mass dark matter halos as `nanolensing.'

In this estimate, we made a significant leap by assuming $\kappa_{\rm back}=0.5$, when in reality it has significant variance.  For a conservative estimate, we can set the background to $\kappa_{\rm back}=0$ and consider the area around a single halo where $\kappa_{\rm 1 halo} \geq 1$.  Then the probabilities drop to 0.01, $7 \times 10^{-5}$, and $3 \times 10^{-12}$ for slopes of $\alpha = 1.5$, $1$, and $0.5$ respectively.  Clearly, for a steep halo profile ($\alpha > 1$) and a sufficient amount of observations, the likelihood of observing a significant nanolensing event is not small.  Subsolar mass dark matter halos with power-law slopes shallower than $\alpha \approx 1$, on the other hand, are unlikely to produce any observable effects.

\section{Micro- and Nanolensing Simulations}
\label{sec:nano}

We create a mock lens system similar to typical observed systems.  The lens galaxy is at $z=0.3$ and the quasar source is at redshift $z=2$.  The magnification of each image in a smooth lens system is given by
\begin{equation}
\mu = \frac{1}{(1-\kappa)^2 - \gamma^2},
\end{equation}
where $\kappa$ is the local projected mass density and $\gamma$ is the external shear.  Images in which $1-\kappa-\gamma>0$ are minima in the time travel surface and are positive parity images, while those with $1-\kappa-\gamma<0$ are saddle points and are negative parity images.  We create a minimum (M) and a saddle point (S) image that in the case for which all the mass is distributed in a smooth component (no stars, no subsolar mass DM halos) have a magnification typical for lens system with four images, $\mu \sim 10$.  The parameters for the mock images are chosen to be the same as in \citet{schechter_wambsganss02} and are shown in Table \ref{tab:sim_param}.

Images in a four image system are located at $\sim$3\% of the projected virial radius of the lensing halo, a distance where a significant fraction of the projected mass is expected to be in stars.  Studies have found a range of values for $\kappa_{*}/\kappa$.  For example, the study of \citet{koopmans_etal06} of the Sloan Lens ACS Survey found an average stellar mass fraction of 0.75 inside of the Einstein radius.  Studies of lens systems with quasar sources have found generally lower stellar mass fractions:  $\geq 0.5$  \citep[Q2237+0305;][]{kochanek04}, 0.08-0.15 \citep[PG 1115+080;][]{morgan_etal08}, 0.1-0.3 \citep[PG 1115+080 and SDSS 0924+0219;][]{schechter_wambsganss04}, $\sim 0.1$  \citep[PG 1115+080;][]{pooley_etal09}, 0.05 \citep{mediavilla_etal09} for a sample of lens systems, and 0.1 \citep[RXJ 1131-1231;][]{dai_etal10}.

We put half of the projected mass in each mock image realization in stars, $\kappa_{*}= 0.5\kappa$, a relatively high value and potentially a large contaminant to the signal from subsolar mass dark matter halos.  Stars are modeled as point mass perturbers with a Salpeter mass function \citep{salpeter55}, $dn/dm \approx m^{-2.35}$ with mass limits of $0.01M_{\sun}$ to $1M_{\sun}$.  Stars are placed at random within an area with radius $\sim20$ times greater than the Einstein radius of a solar mass star.  

We introduce subsolar mass halos in the simulations in the following way. If subsolar mass dark matter halos are resistant to tidal disruption, they could make up the majority of the dark matter halo of the Galaxy.  The effects of tidal disruption on subsolar mass halos are still under debate, though recent studies suggest that the central cores will survive all but the densest regions ($\leq 20$pc from the Galactic center) of the Milky Way  \citep{ishiyama_etal10}.  Based on this result and for the sake of simplicity, we put the remaining half of the projected mass into subsolar mass halos. We model subsolar mass dark matter halos with power-law density profiles where $\Sigma(R) \sim R^{-1.5}$.  As discussed earlier, this may be steeper than expected and makes them more effective perturbers.   A solar mass dark matter halo has a radius of 1 parsec.  For smaller halos, the radius is scaled with the mass, $M \propto r^{3}$.  The subsolar mass halo mass function is modeled as $dn/dm \approx m^{-2}$ with mass range of $10^{-4}M_{\sun}$ to $1M_{\sun}$.  These halos are distributed randomly within the same area as stars.  While stars, as point masses, always induce microlensing as long as light from the source passes close enough to the star, subsolar mass halos may not act alone.  Each realization has significantly more subhalos than stars ($\sim10^5$ compared to $\sim10^3$), and many halos overlap.  As a result, the mass density from every halo adds up and microlensing events are induced where the combined mass density of many subhalos exceeds the critical density for lensing.  
%
\begin{table}
\begin{center}
\begin{tabular}{c|c|c|c}
Mock Image & $ \kappa $ & $ \gamma $ & $ \mu $  \\
\hline
Minimum (M) & 0.475 & 0.425 & 10.5 \\
Saddle (S)  & 0.525 & 0.575 & -9.5 \\
\end{tabular} 
\caption{Simulation parameters}
\label{tab:sim_param}
\end{center}
\end{table}

We generate seven realizations for each mock saddle image (labeled Sh1 to Sh7), and five realizations for each mock minimum image (Mh1 to Mh5).  For comparison, we also generate the same number of realizations containing only stars but no subsolar mass dark matter halos -- in this case all the dark matter is in a smooth component (S1 to S7 and M1 to M5).

We perform a lensing simulation using an inverse ray-tracing code.  The position of the source, $\vec{\beta}$, and the images, $\vec{\theta}$, are related by the lens equation, $\vec{\beta} = \vec{\theta} - \nabla \phi(\vec{\theta})$, where $\phi$ is the lensing potential. In the case of circular symmetry in the lens, $\nabla \phi \propto M$, where $M$ is the mass of the lens.  As the lens equation is multi-valued and several image positions may correspond to a single source position, it is easiest to solve by establishing a grid of image positions and calculating the source position for each point on the grid.  Every set of three points in the image plane defines a triangle and corresponds to a triangle in the source plane.  The ratio of the area of the image plane triangle to the source plane triangle defines the magnification of that area of the source plane.  

\begin{figure*}[t]
\epsscale{1.15}
\centering
\plottwo{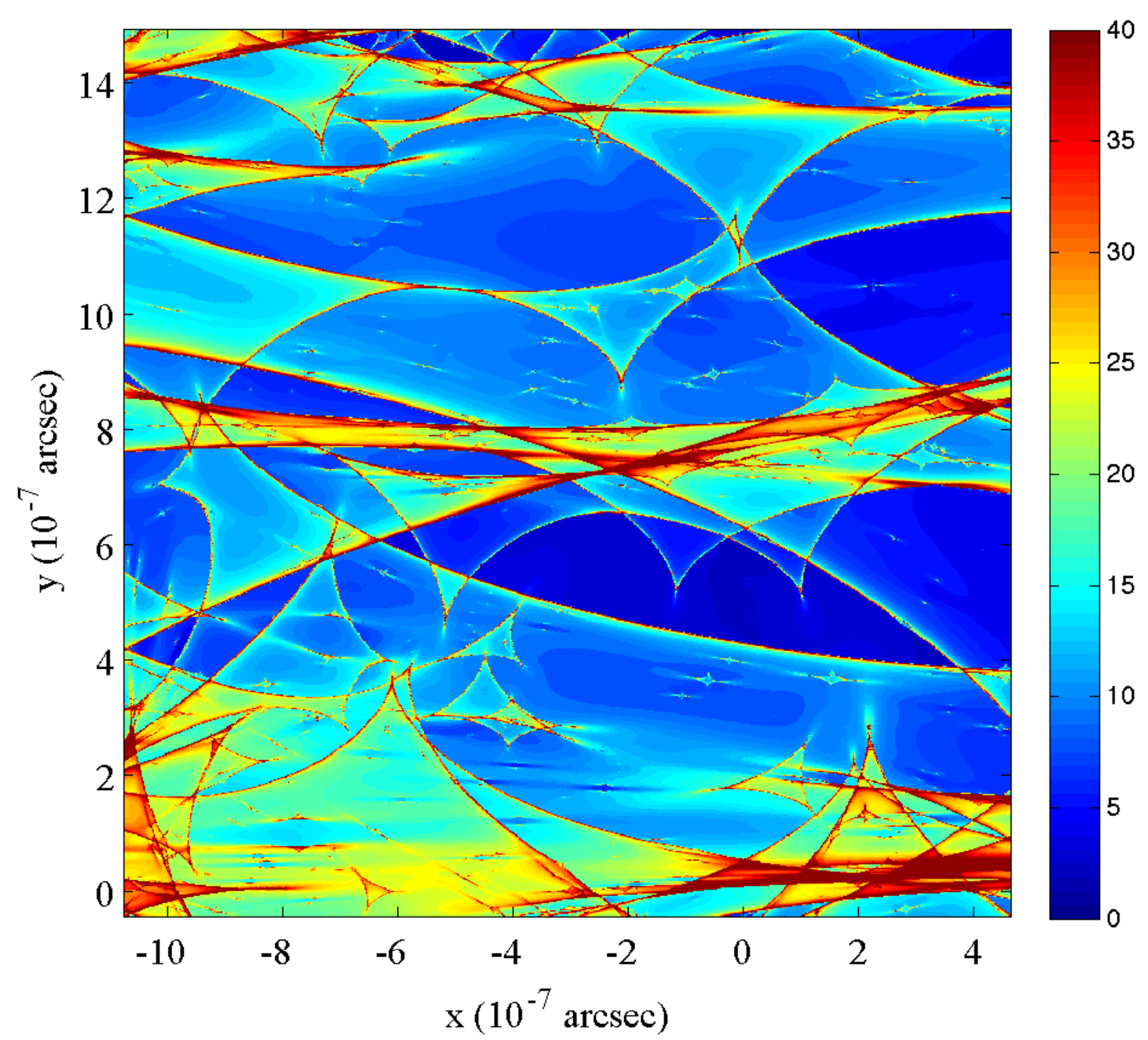}{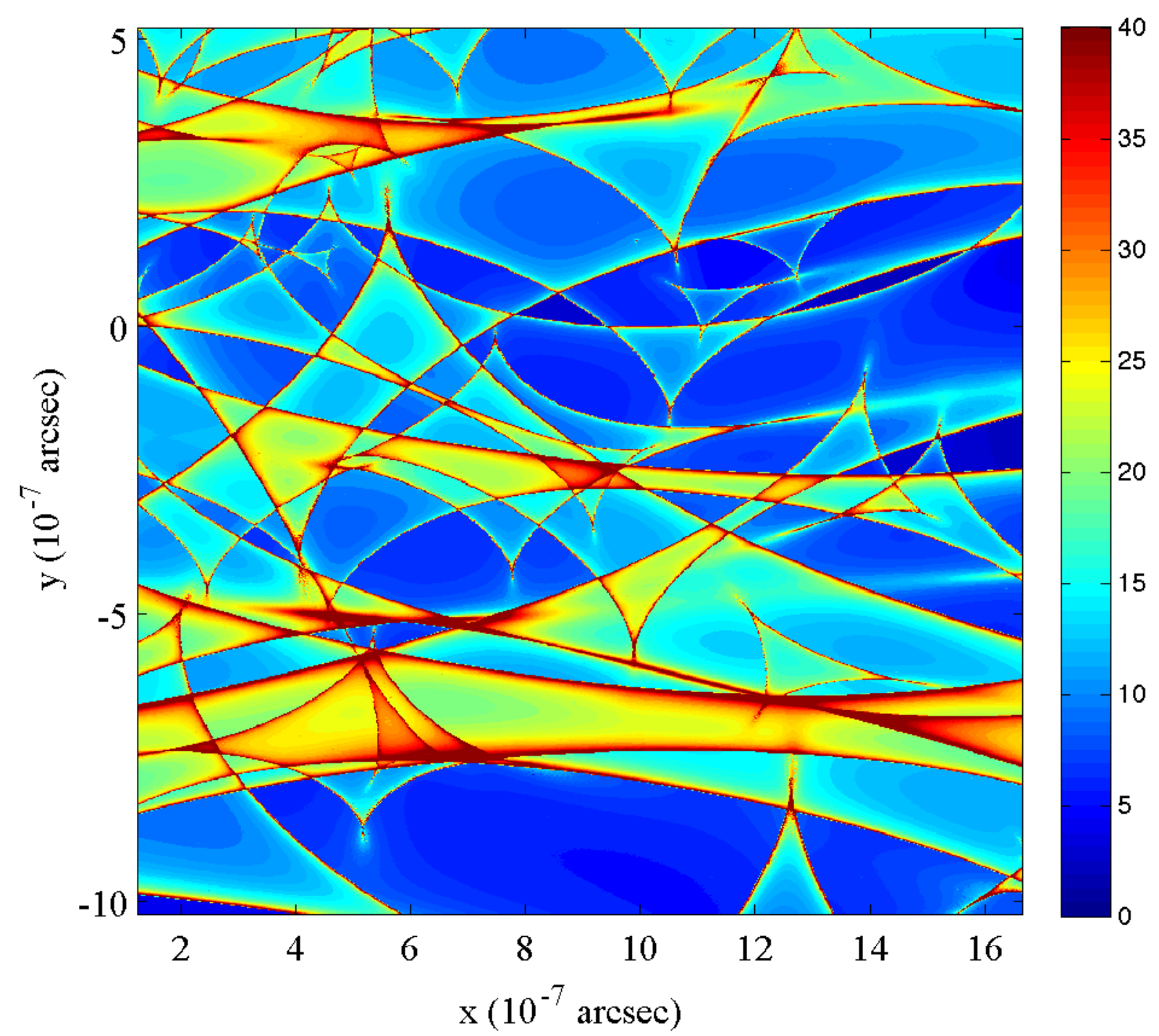}
\plottwo{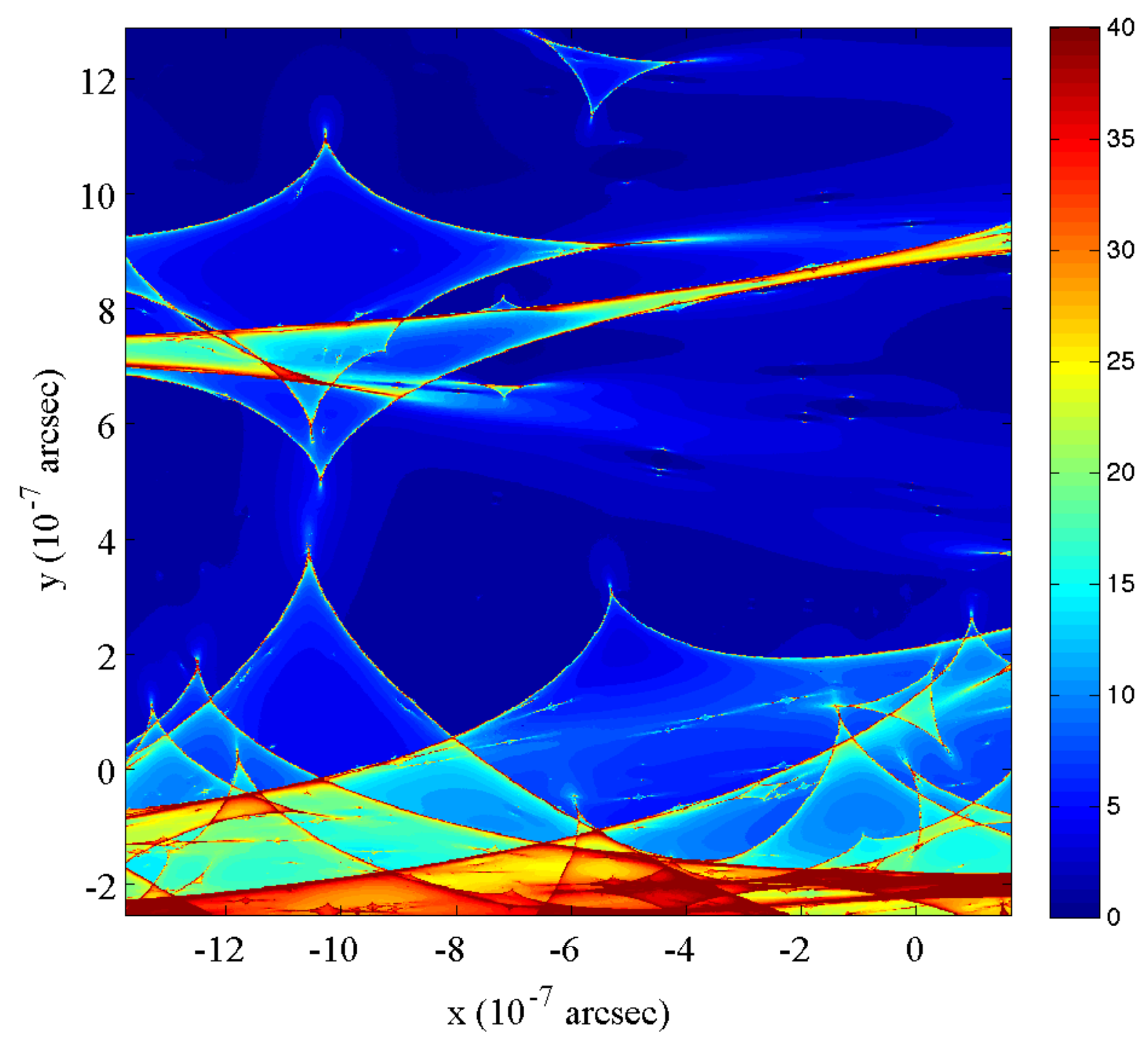}{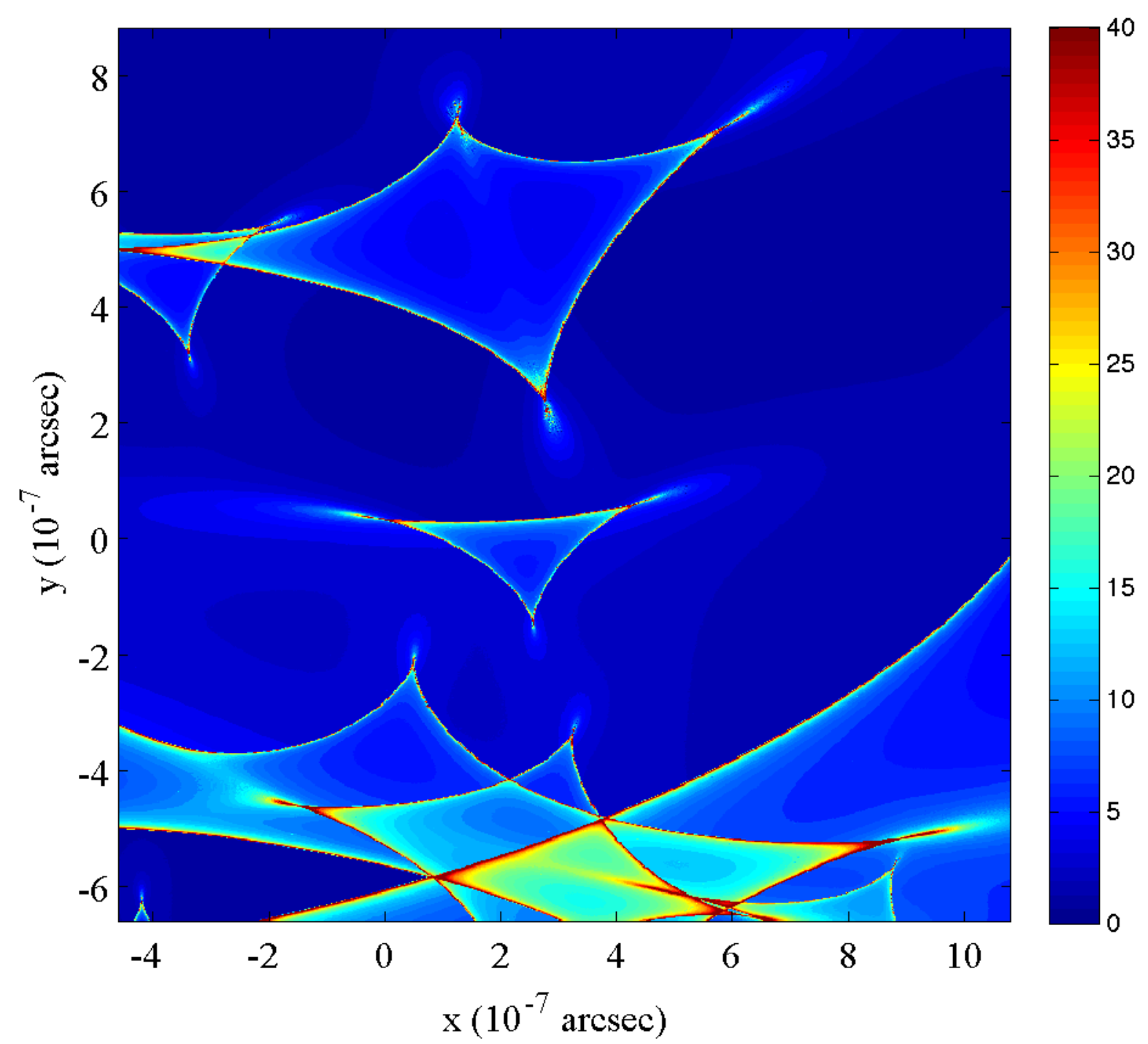}
\caption{Sample magnification maps for a saddle point image.  The realizations on the left  (Sh1, {\it top};  Sh2, {\it bottom}) contain subsolar mass halos, while those on the right (S1, {\it top};  S2, {\it bottom}) are without.  Note that the realizations to the left have more fine structure than the realizations to the right.  \label{fig:magmap_saddle}}
\end{figure*}

\begin{figure*}[t]
\epsscale{1.15}
\centering
\plottwo{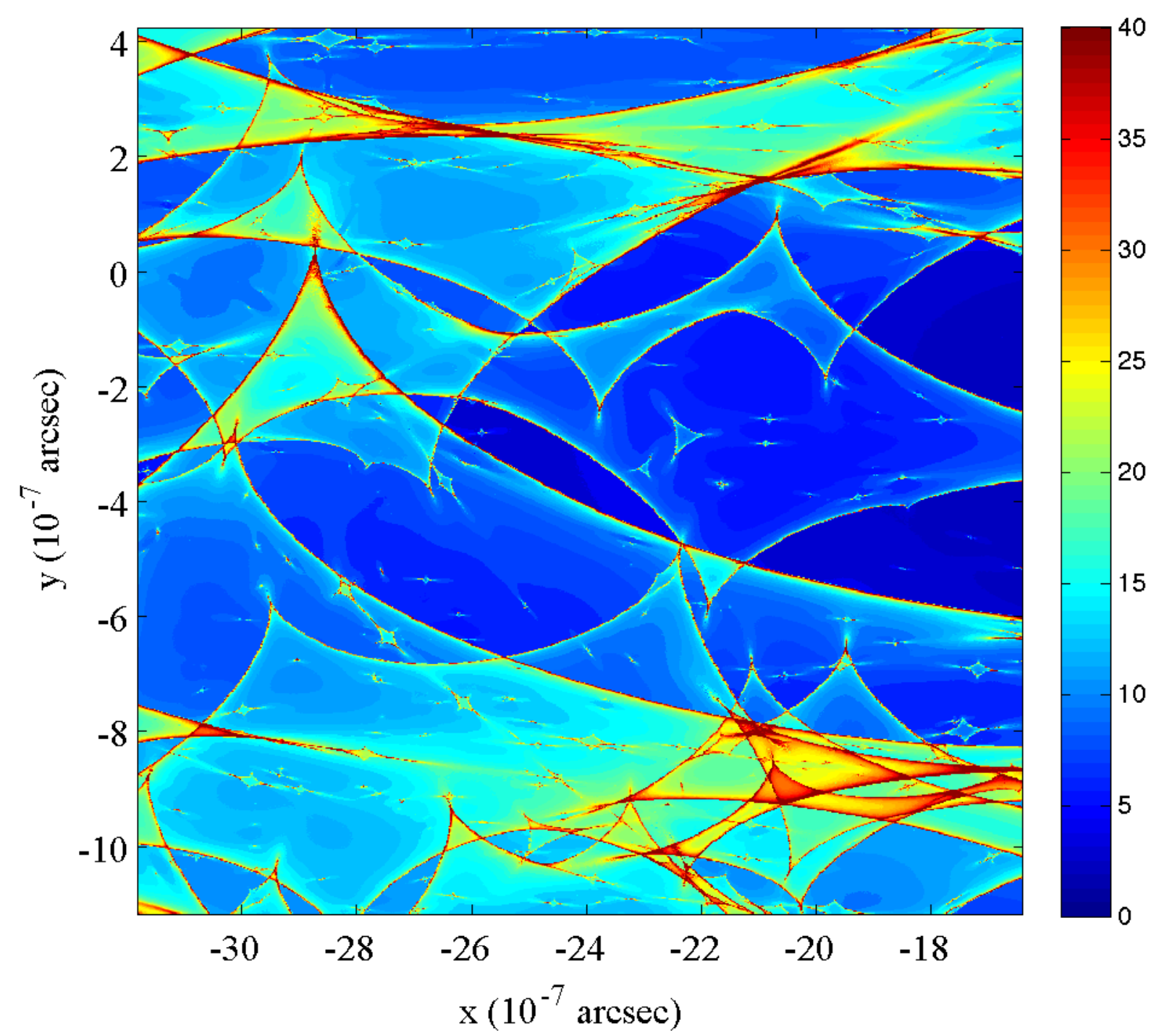}{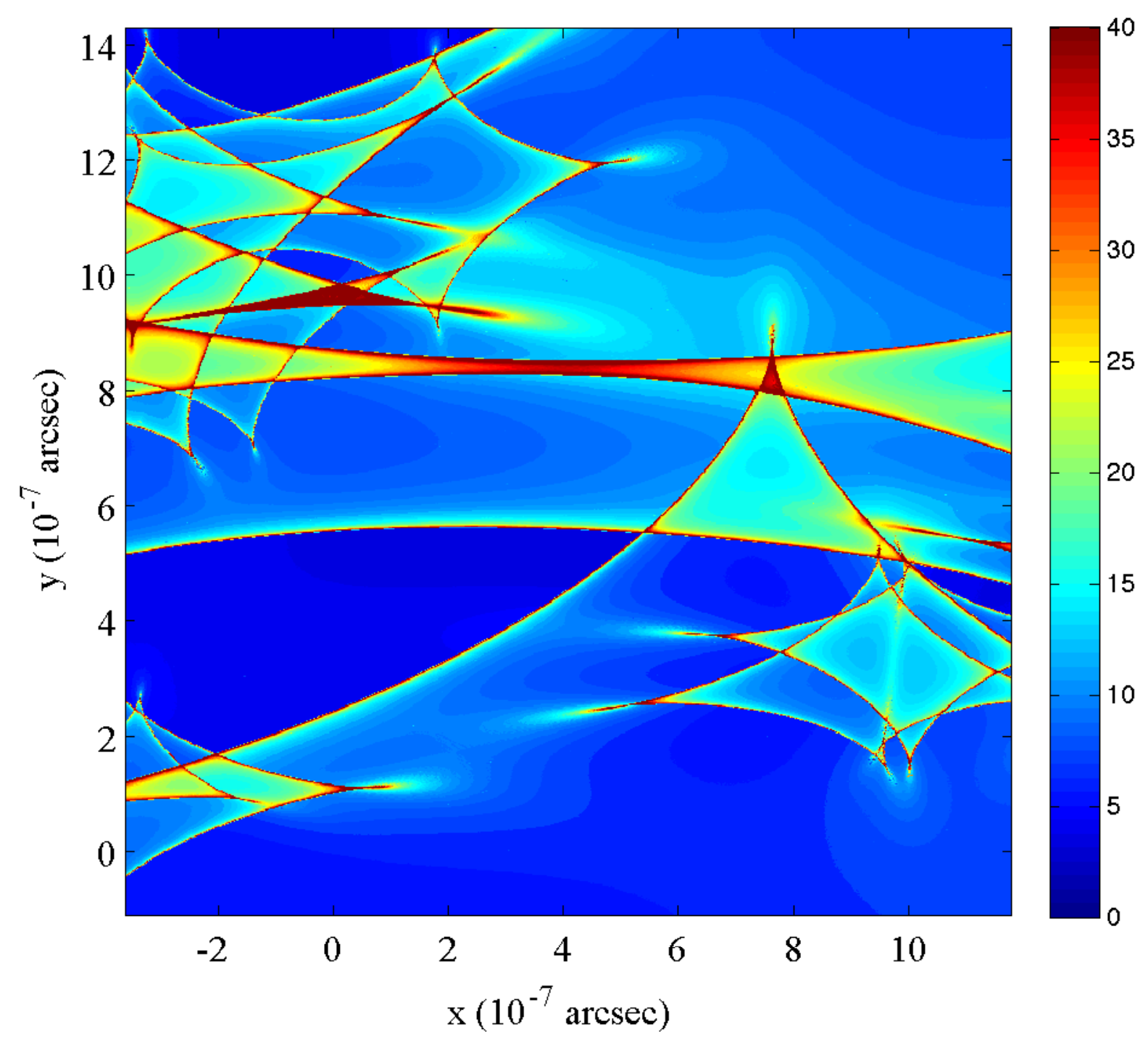}
\plottwo{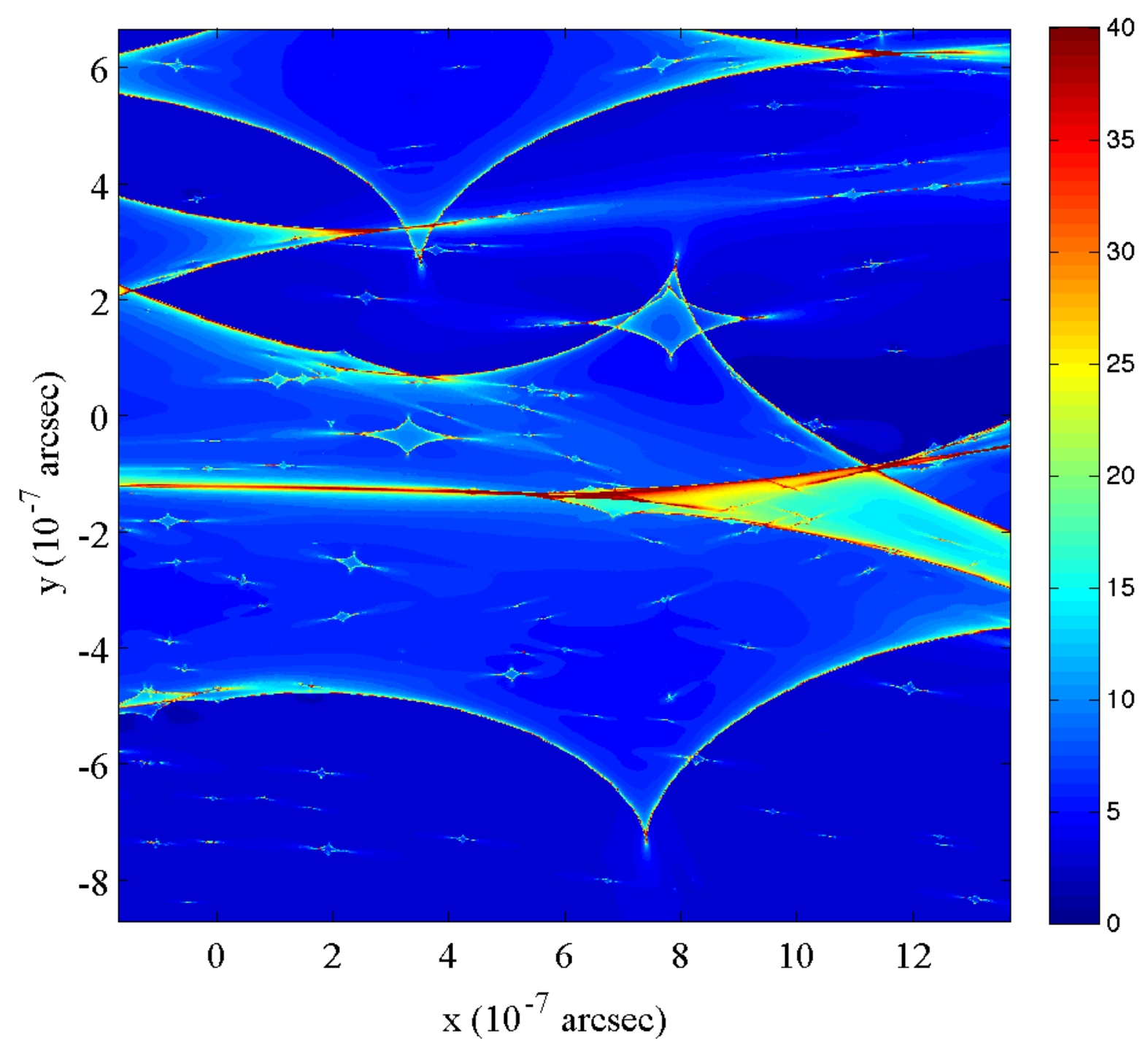}{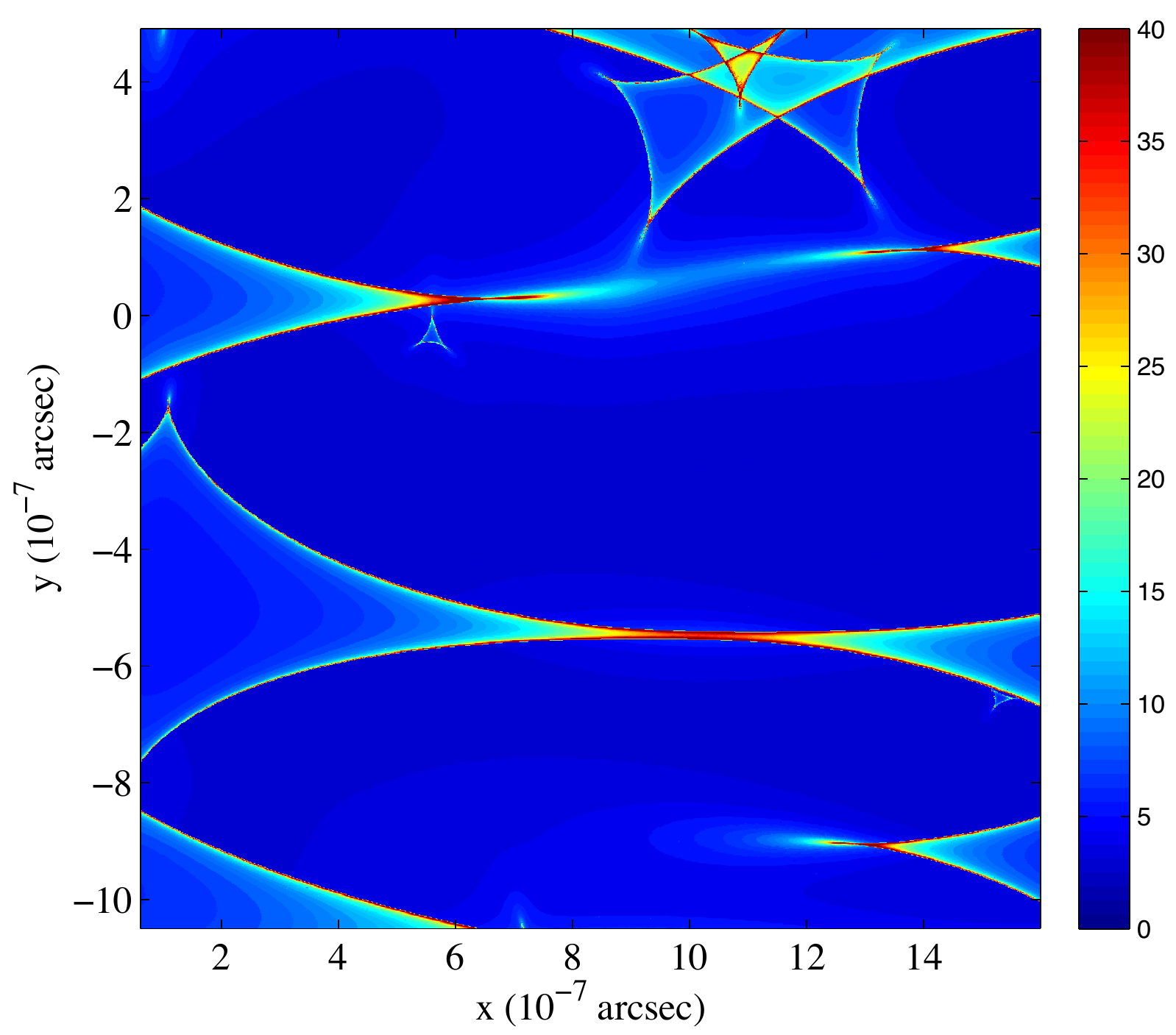}
\caption{Sample magnificatixon maps for a minimum point image.  The realizations on the left  (Mh1, {\it top};  Mh2, {\it bottom}) contain subsolar mass halos, while those on the right (M1, {\it top};  M2, {\it bottom}) are without.  Note that the realizations to the left have more fine structure than the realizations to the right.  \label{fig:magmap_min}}
\end{figure*}

In Figs.~\ref{fig:magmap_saddle} \&~\ref{fig:magmap_min} we show the magnification maps for four realizations of minimum and four realizations of saddle points.  Each map shows a small patch of the source plane, the color corresponding to the total magnification of the image (including any micro- or nanoimages) for a source at that position.  Given the small size of the patches and the stochasticity of the placement of the stars and halos, different realizations with the same properties look significantly different.  

As previous works have also shown \citep[e.g.,][]{schechter_wambsganss02}, the magnification maps show a wealth of features that can be described as of two types:  peaks (or cusps) of high magnification, and valleys of lower magnification.  In particular, the saddle points -- which are preferentially demagnified by perturbations -- show large valleys of very low magnification.  In comparison to the maps that do not contain subhalos, maps with subsolar mass halos show a significant excess of small structures, and the saddle point realizations have several smaller, even lower magnification, valleys layered on top of the larger valleys.

\section{Mock Light Curves}
\label{sec:mock}

The magnification maps offer a detailed look into the lensing effects of subsolar mass dark matter halos.  Observations, however, are limited to sampling a light curve as the source moves across the magnification map.  A single light curve will contain a small fraction of the information in the magnification map.  Any distinct feature, such as a valley, that is observed in a track crossing in one direction will look very different if the source were to move in a different direction.  In addition, finite sources will smear out some features.  Finally, the observable is fluxes -- and not magnifications -- of images.  Connecting observed fluxes to predicted magnifications require a detailed understanding of the smooth lens component, the lens environment, and the time delay between images.

We create a catalog of mock light curves from the realizations of magnification maps.  The effective source-plane velocity is 
\begin{equation}
\vec{v}_{e} = \frac{\vec{v}_{o}}{1 + z_l} \frac{D_{ls}}{D_{ol}} - \frac{\vec{v}_{l}}{1+z_{l}}\frac{D_{os}}{D_{ol}} + \frac{\vec{v}_{s}}{1+z_s},
\end{equation}
where $\vec{v}_o$, $\vec{v}_l$, and $\vec{v}_s$ are the velocities of the observer, the lens, and the source;  $z_l$ and $z_s$ are the redshifts of the lens and the source; and $D_{ol}$, $D_{ls}$, and $D_{os}$ are the angular diameter distances of the observer to the source, the lens to the source, and the observer to the source.  We set a fiducial effective velocity of 600 km s$^{-1}$.  

The source is modeled as a uniform disk, with fiducial radius of $2 \times 10^{14}$ cm.
While the source profile is simplistic, \citet{mortonson_etal05} show that the microlensing signal is mostly sensitive to the half-light radius of the disk and the details of the disk profile are less important.  The size of the quasar source depends on the wavelength of the observation.  The smallest sources are observations of optical continuum flux, $\sim10^{15}$ cm \citep[see, e.g.,][]{morgan_etal10}.  Our fiducial disk size is smaller than previously measured values but falls near the range of values expected and is larger than the radius of the last stable orbit for a Schwarzschild black hole of $2 \times 10^{8} M_{\sun}$. 

Each magnification map is $1.5 \times 10^{-6}$ arcsec on a side.  At the fiducial velocity, an interval of 10 years will traverse an angular distance of $7 \times 10^{-7}$ arcsec.  Therefore, based on the size of the magnification maps, we can create light curves that contains 10 years of data sampled at week intervals.  During that time interval the source moves in a trajectory that is randomly oriented with respect to the magnification map.  

\begin{figure}[t]
\epsscale{1.15}
\plotone{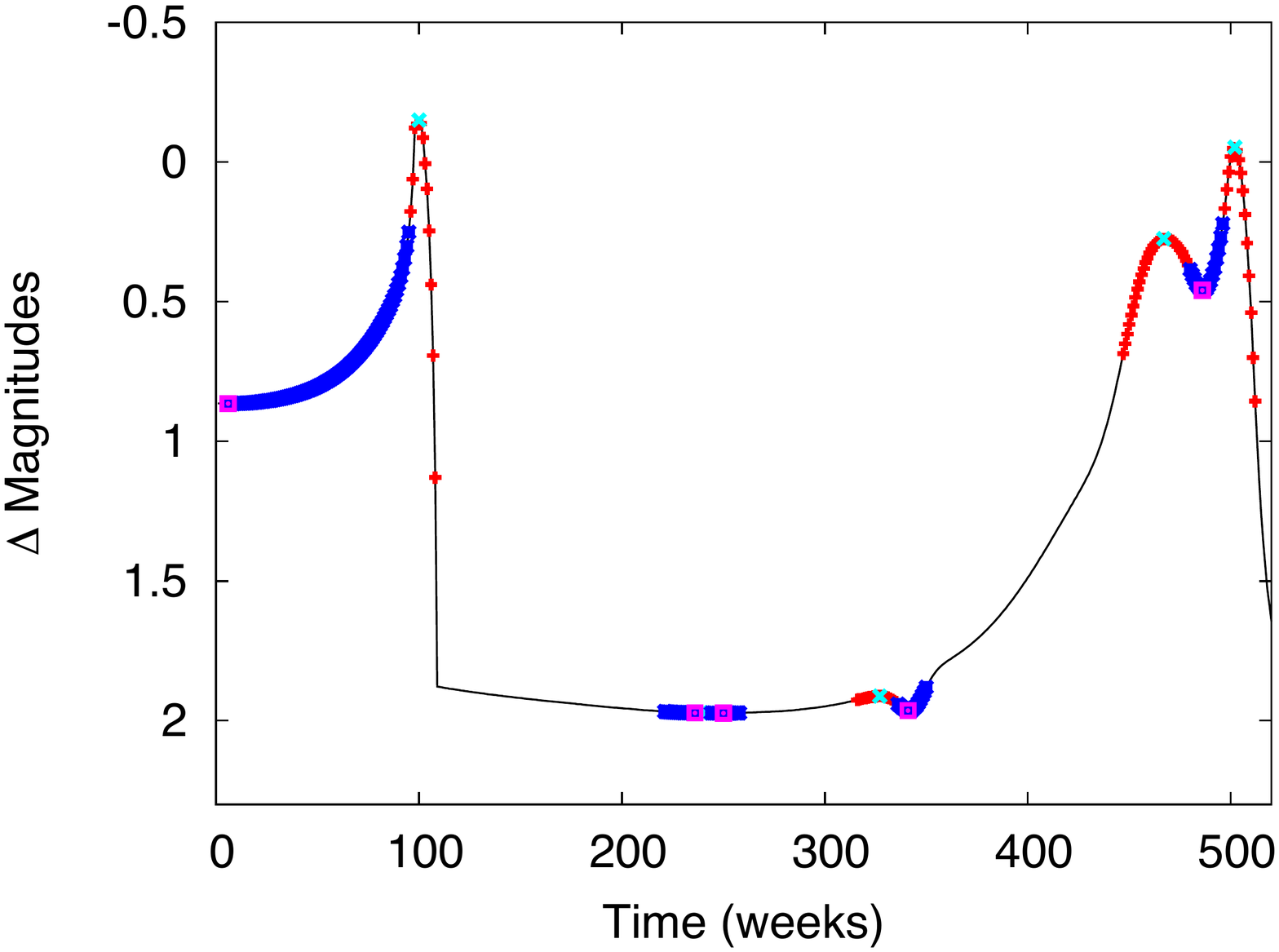}
\plotone{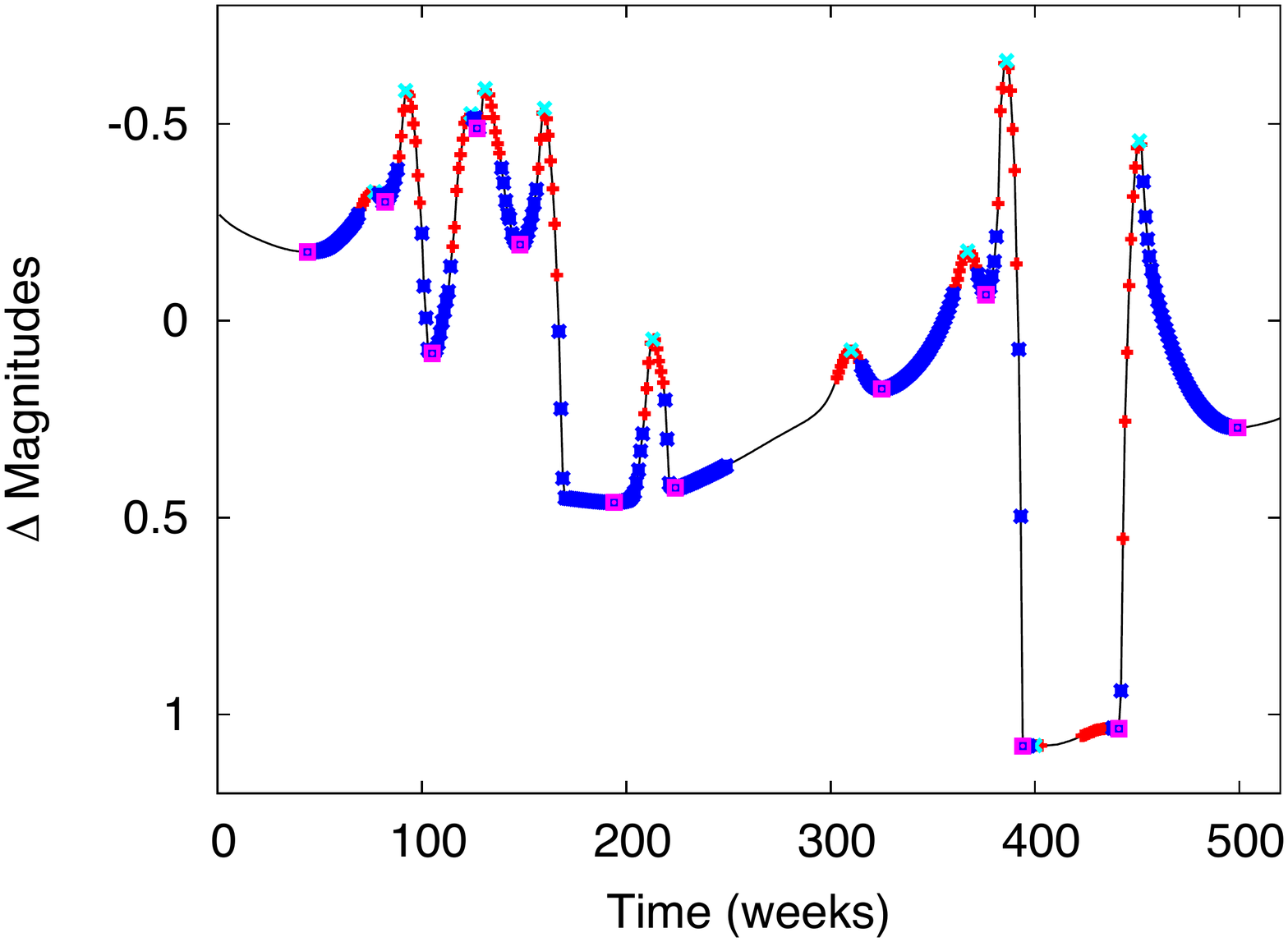}
\caption{Sample 10-year light curve with observations taken weekly drawn from Sh2 {\it (top)} and from Mh1 {\it (bottom)}.  Peaks are labeled in red points and valleys are labeled in blue points.  The extrema for those events are labeled in cyan and magenta points, respectively.  Portions of the light curves remain unlabeled;  they are referred to as `plains,' although they may be asysmptotically rising or falling.  The effects of subolar mass halos can be seen as a small valley around week $\sim 350$ in the top panel and the three small peaks between weeks 175 and 375 in the bottom panel.
\label{fig:example_lc}}
\end{figure}

We create 200 such light curves for each realization.  Figure \ref{fig:example_lc} shows sample light curves.  Here some typical features are readily apparent:  long planes of low magnification and cusps of high magnification at caustic crossings.  In addition, some features can be identified as caused by subsolar mass halos:  a small valley at week $\sim 350$ in the upper panel and several small peaks between week 175 and week 375 in the lower panel.  

\begin{figure}
\epsscale{1.15}
\plottwo{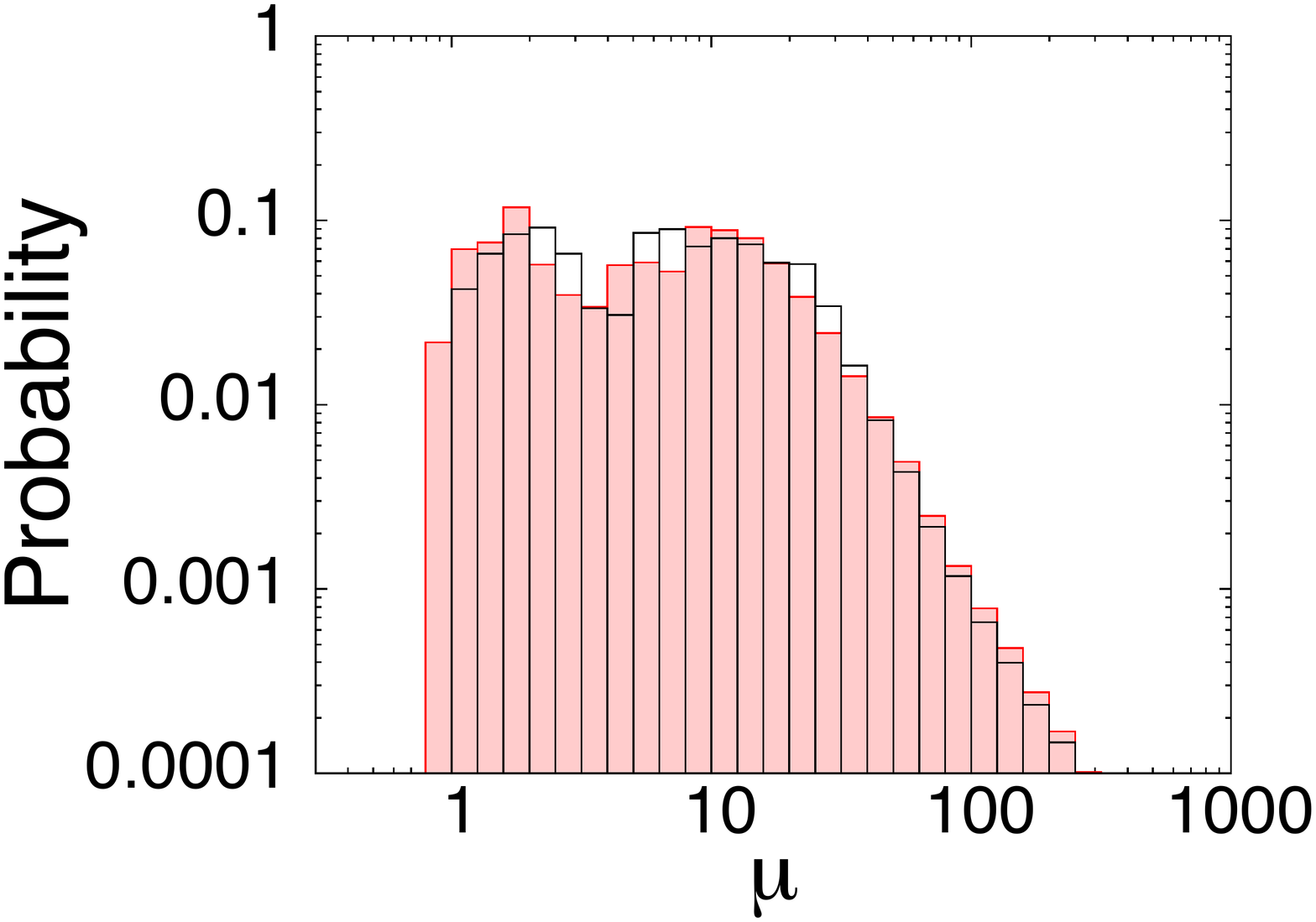}{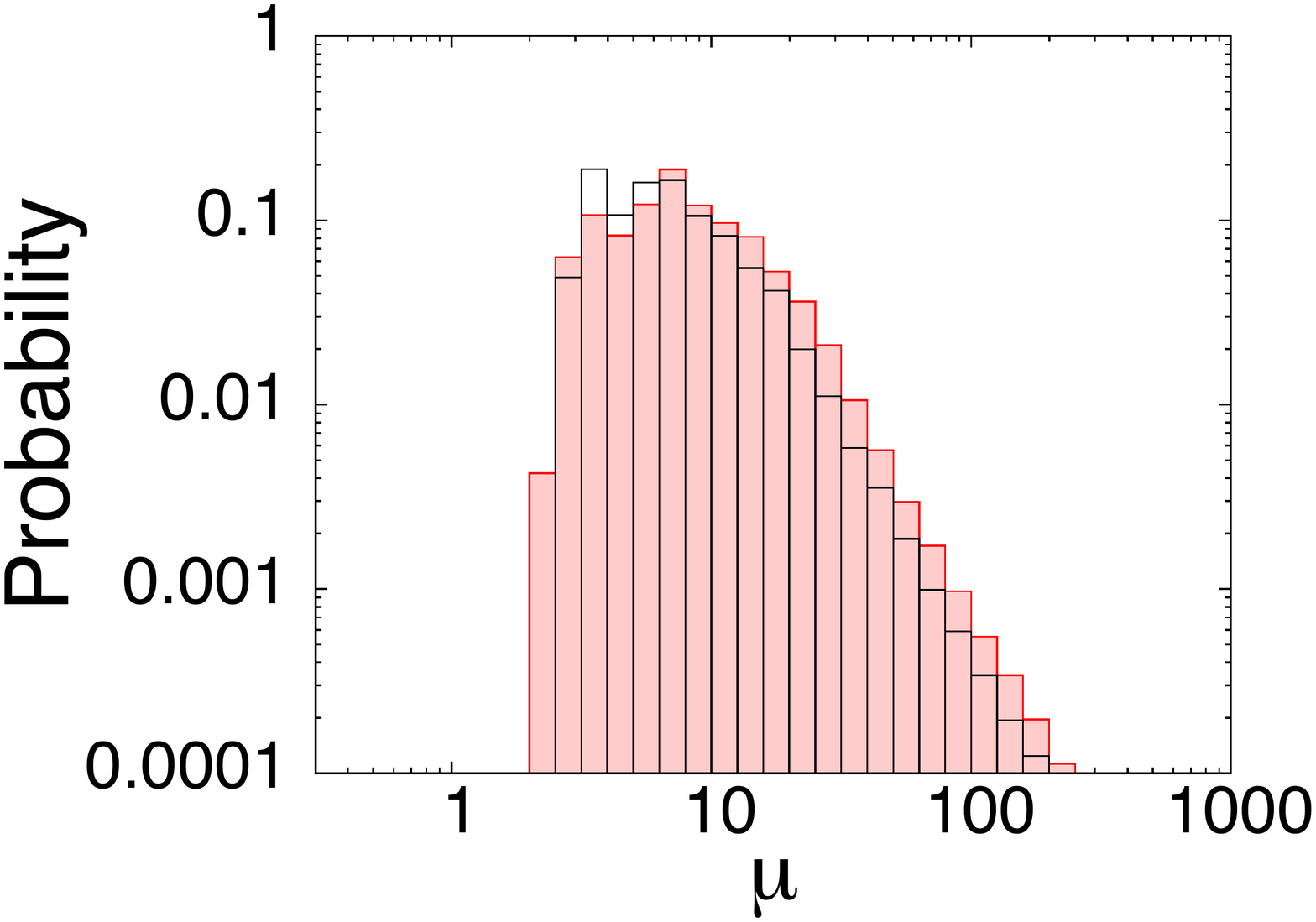}
\caption{Magnification histogram for all saddle point realizations {\it (left)} and all minimum point realizations {\it (right)}.  The filled red bars represent the realizations with subsolar mass halos, while the black lines represent the realizations without. Simple magnification histograms are unable to distinguish between realizations with subsolar mass halos and those without.
\label{fig:mag_histo}}
\end{figure}

Histograms of the magnification maps are shown in Figure \ref{fig:mag_histo}.  Any differences in the histograms between realizations with subsolar mass halos from those without is small:  a slight excess of lower magnifications both saddle and minimum points, a slight excess of higher magnifications in the minimum points.  Without knowing the magnifications (and not the observed fluxes) of the images, these small differences may be impossible to measure at all.  It is clear, then, that a simple statistical analysis of the measured fluxes in observed light curves will be similarly insufficient in detecting subsolar mass DM halos.  We must, instead, seek an analysis of the {\it features} of the light curves in order to find more significant differences between realizations with dark matter halos and those without.

\section{Analyzing Light Curve Features}
\label{sec:analysis}

Our mock light curves are without observational errors, and so a simple analysis can be utilized.  We find the centers of peaks and valleys by the derivative of the light curve.  The beginning and the end of a peak or valley is found by the curve's inflection points.  Events that extend beyond the length of the light curve are ignored.  Peaks and valleys which are not at least $\Delta$ mag $=0.01$ are discarded.  $\Delta$ mag $=0.01$ is significantly larger than the precision of the lensing simulation and similar to that expected from future observations. Accounting for all peak and valley events, some fraction of each light curve will lie between events.  These areas will be referred to as `plains,' and they may be asymptotically rising or falling or flat.  Figure \ref{fig:example_lc} labels the peaks and valleys on a sample light curve.

Table \ref{tab:events} shows the results of the analysis.  The number of events is similar for peaks, valleys, and plains, although the realizations with dark matter halos have $\sim$50-100\% more events  than those without.  

A more in depth look into the characteristics of the events considers the magnitude displacement and timescale of events.  A peak or valley event consists of a rising side, an extremum, and a falling side.  We measure two magnitude amplitudes:  one between the start of an event and the extremum and one between end of the event and the extremum.  We sort the magnitude amplitudes by size -- each peak has a minimum magnitude difference and a maximum magnitude difference.  For plains, the magnitude amplitude is measured as the difference between the start and the end of the event.  The timescales are measured as the total duration of an event.

\begin{figure*}[t]
\centering
\resizebox{2.3in}{!}	{\includegraphics{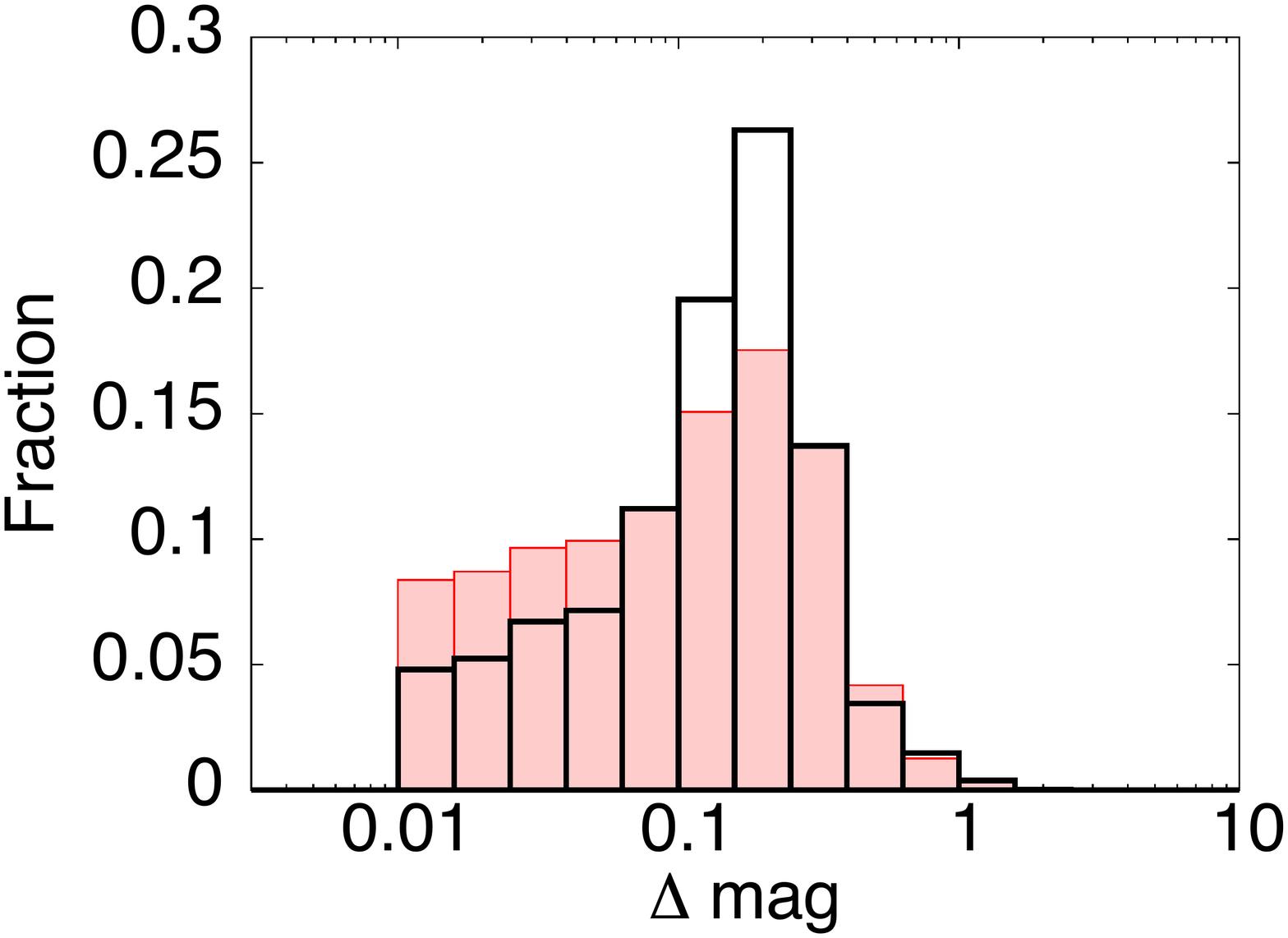}}
\resizebox{2.3in}{!}	{\includegraphics{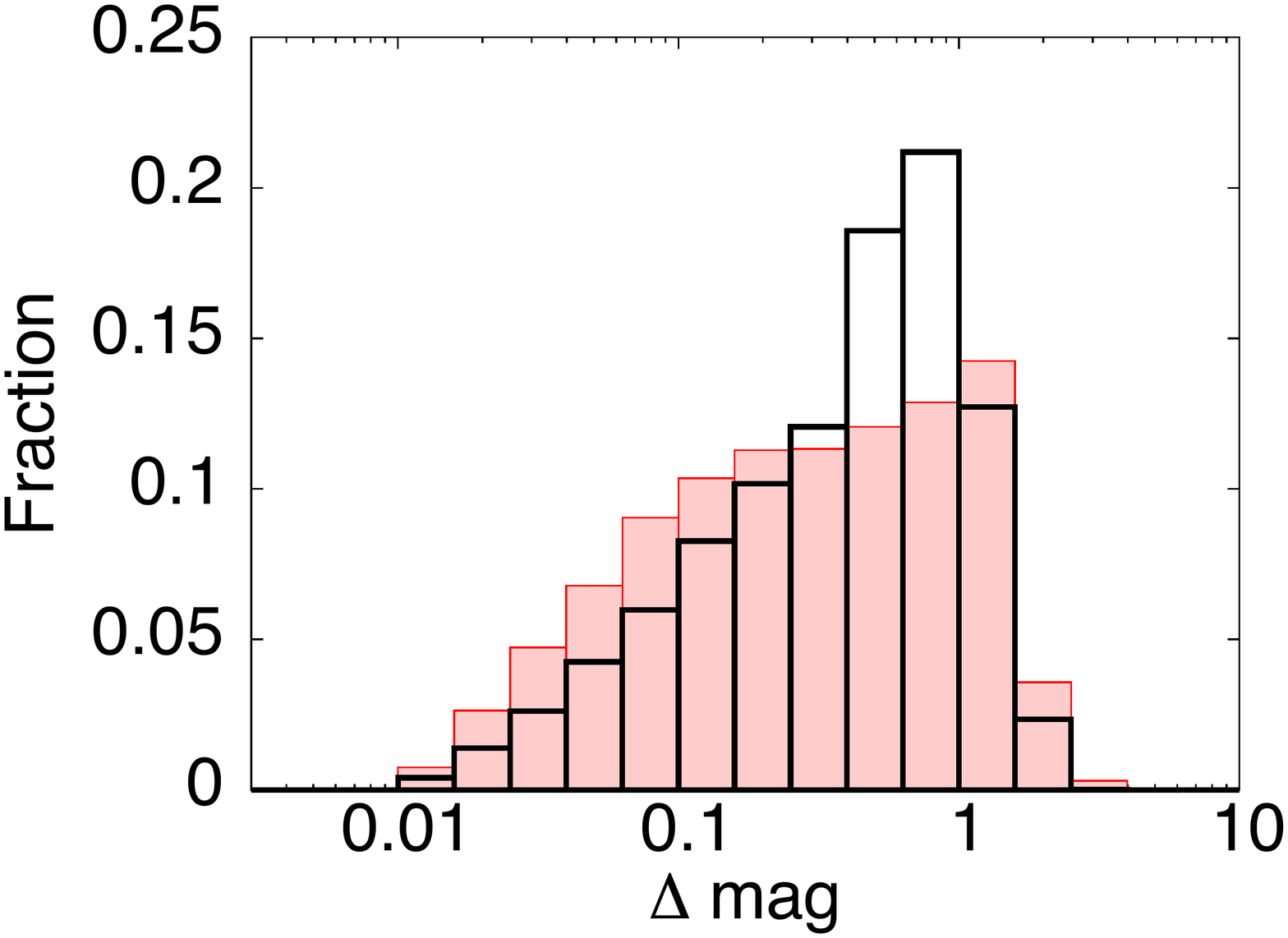}}
\resizebox{2.3in}{!}	{\includegraphics{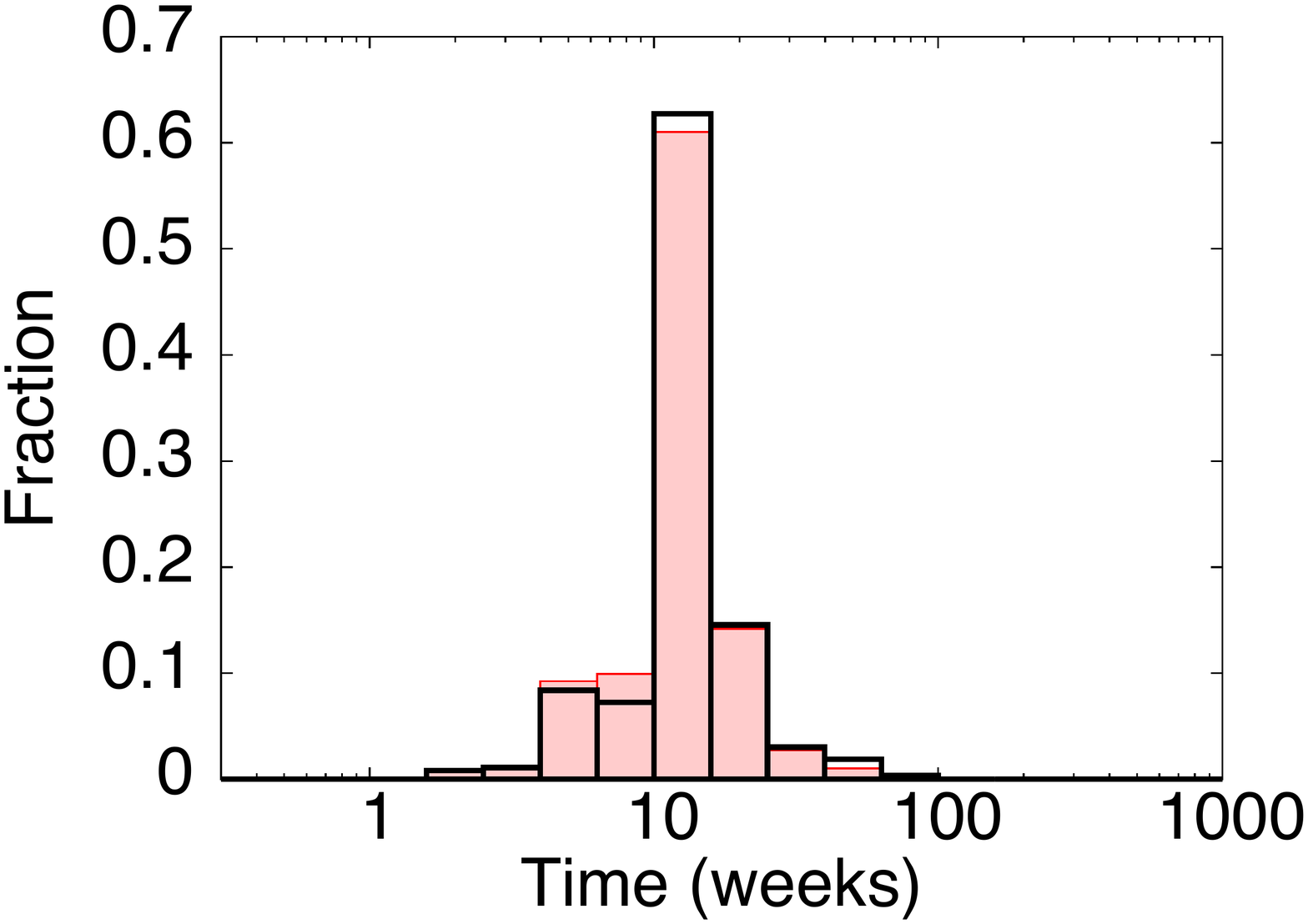}}
\resizebox{2.3in}{!}	{\includegraphics{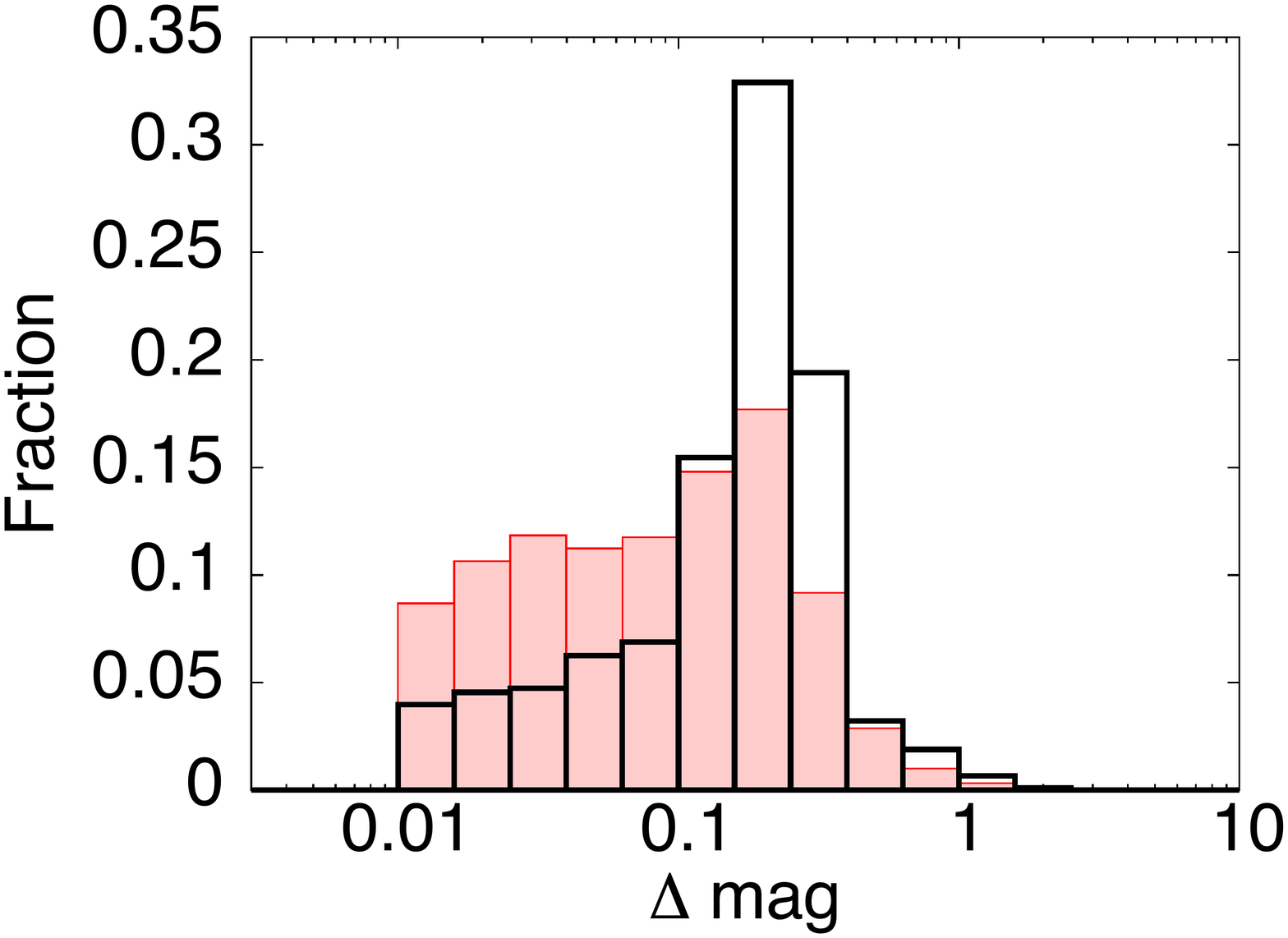}}
\resizebox{2.3in}{!}	{\includegraphics{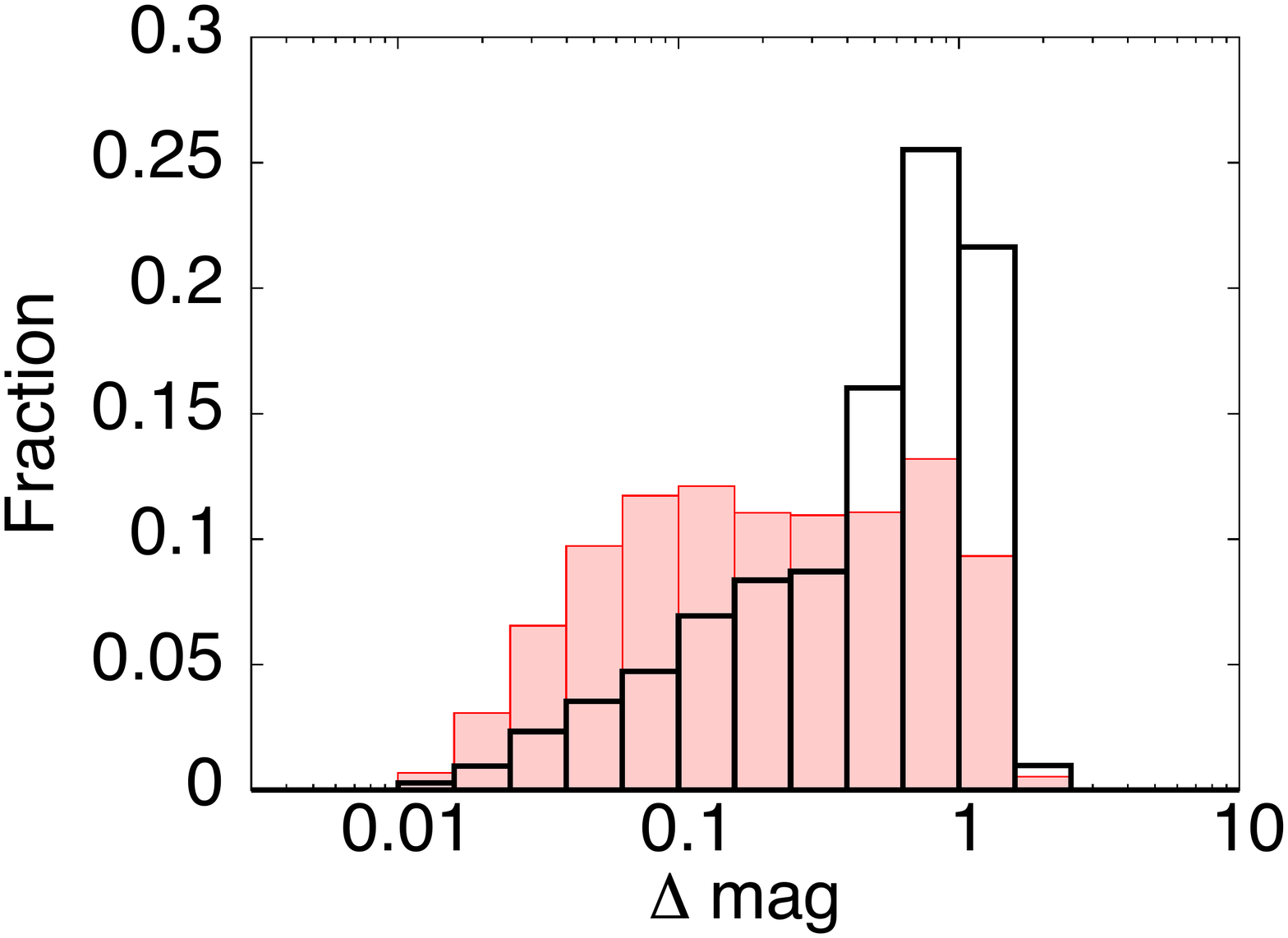}}
\resizebox{2.3in}{!}	{\includegraphics{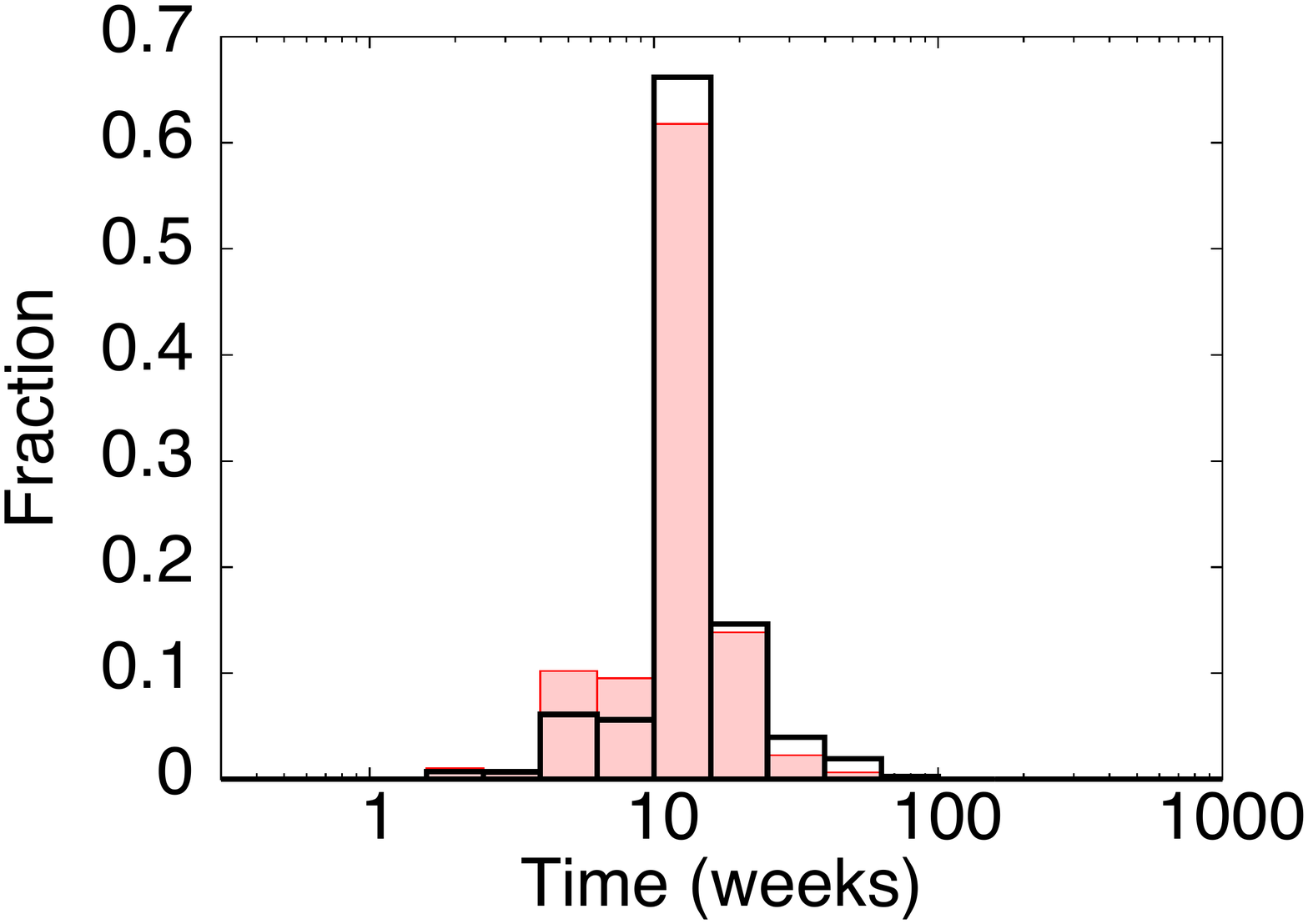}}
\caption{Statistics for `peaks' in saddle point realizations {\it (top)} and minimum point realizations {\it (bottom)}.  Each event consists of a rising side, a maximum, and a falling side.  {\it Left:}  The smaller change in magnitude over the event (either the rising side or the falling side).  {\it Center:}  The larger change in magnitude.  {\it Right:}  The duration of the event in weeks.  The filled red bars represent the realizations with subsolar mass halos, while the black lines represent the realizations without.  \label{fig:lc_peak}}
\end{figure*}

\begin{figure*}[t]
\centering
\resizebox{2.3in}{!}
	{\includegraphics{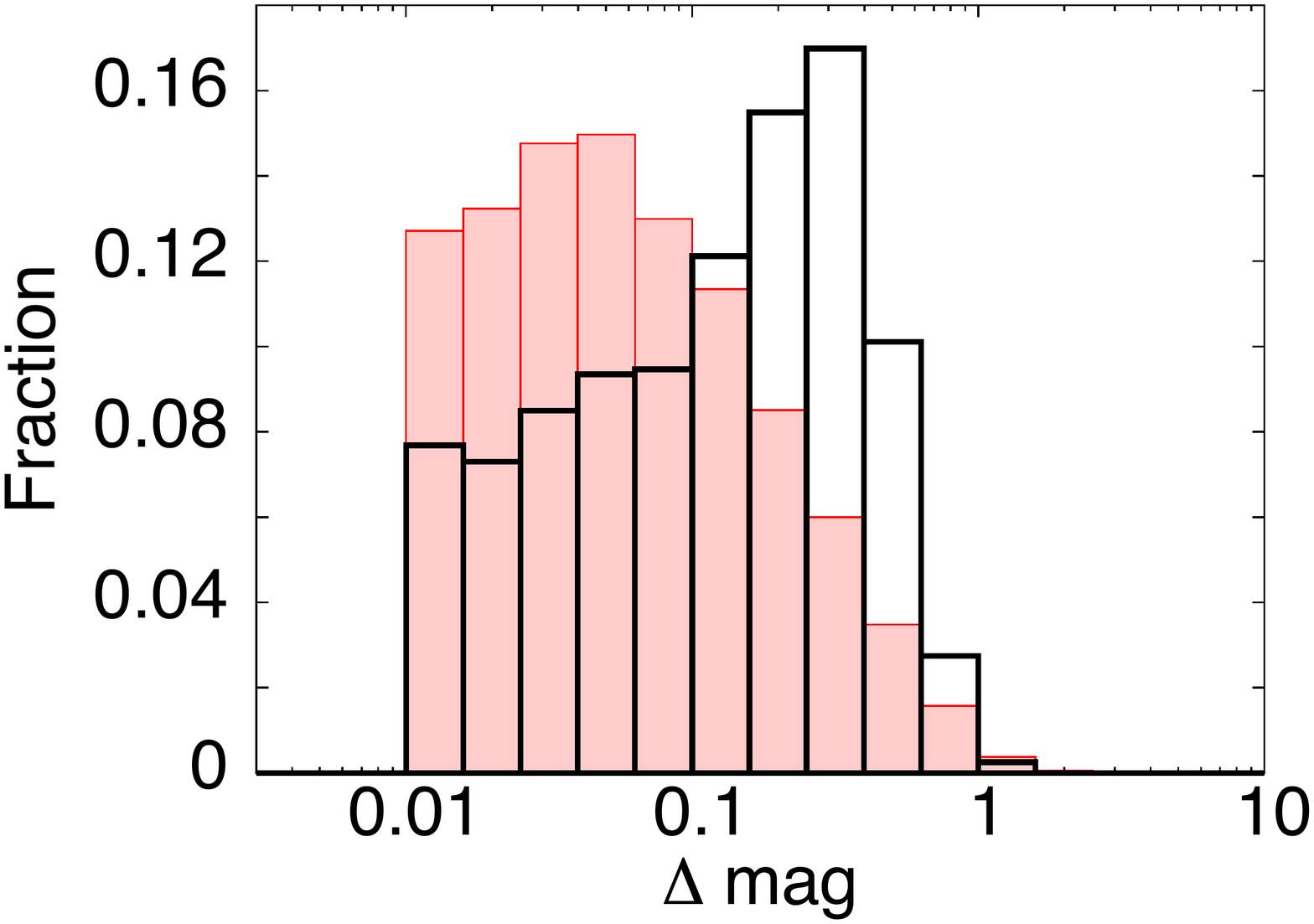}}
\resizebox{2.3in}{!}
	{\includegraphics{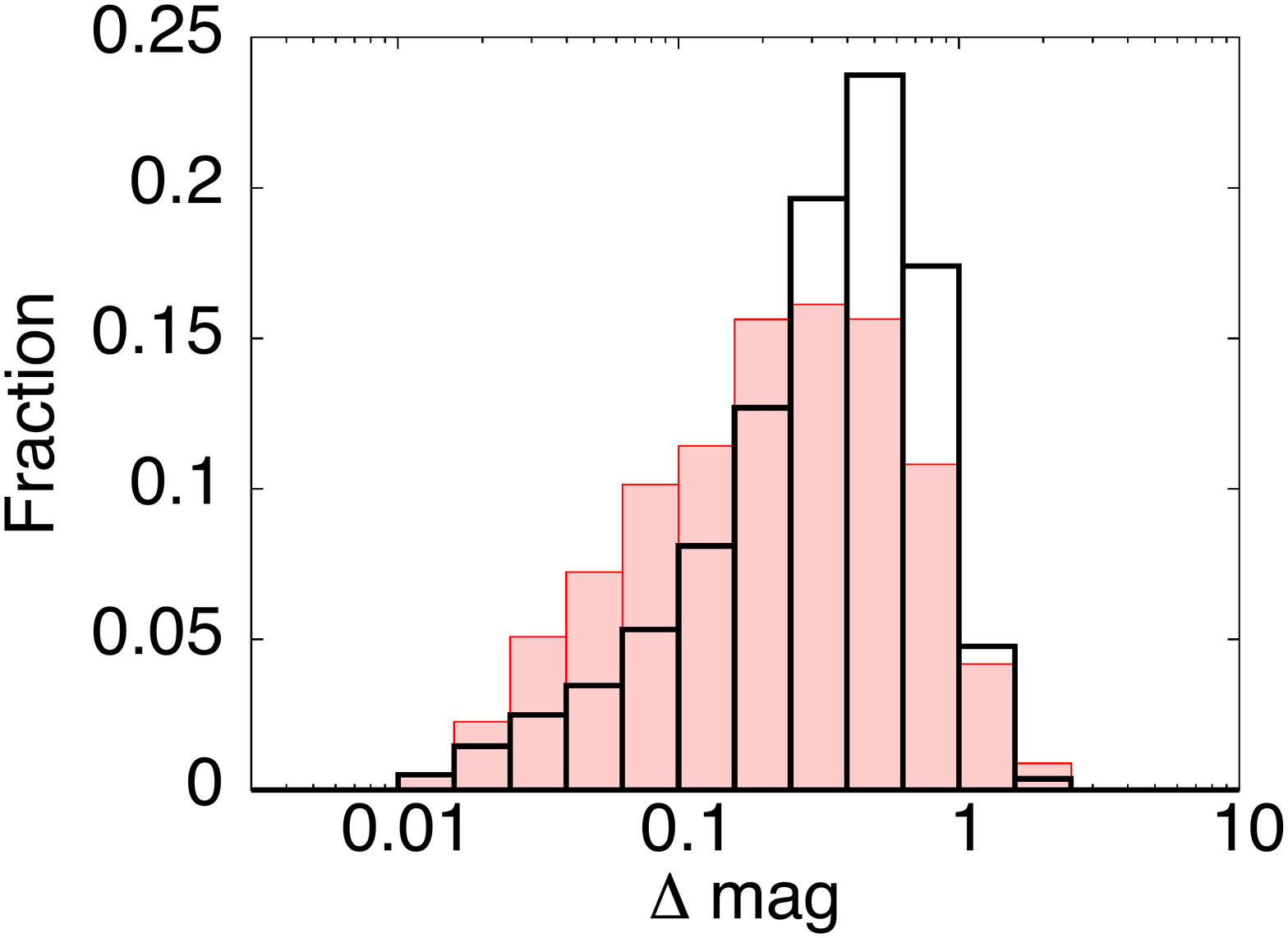}}
\resizebox{2.3in}{!}
	{\includegraphics{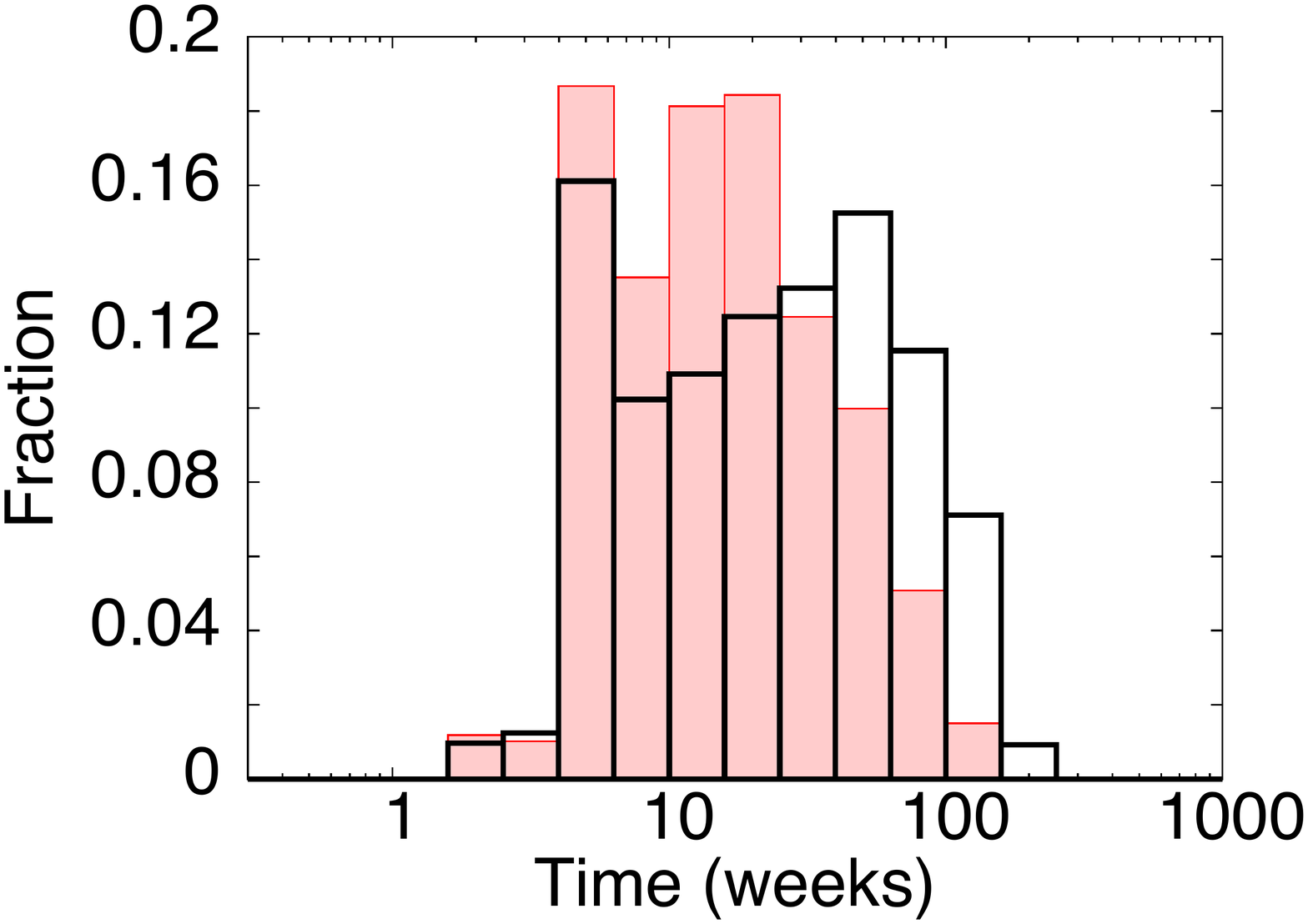}}
\resizebox{2.3in}{!}
	{\includegraphics{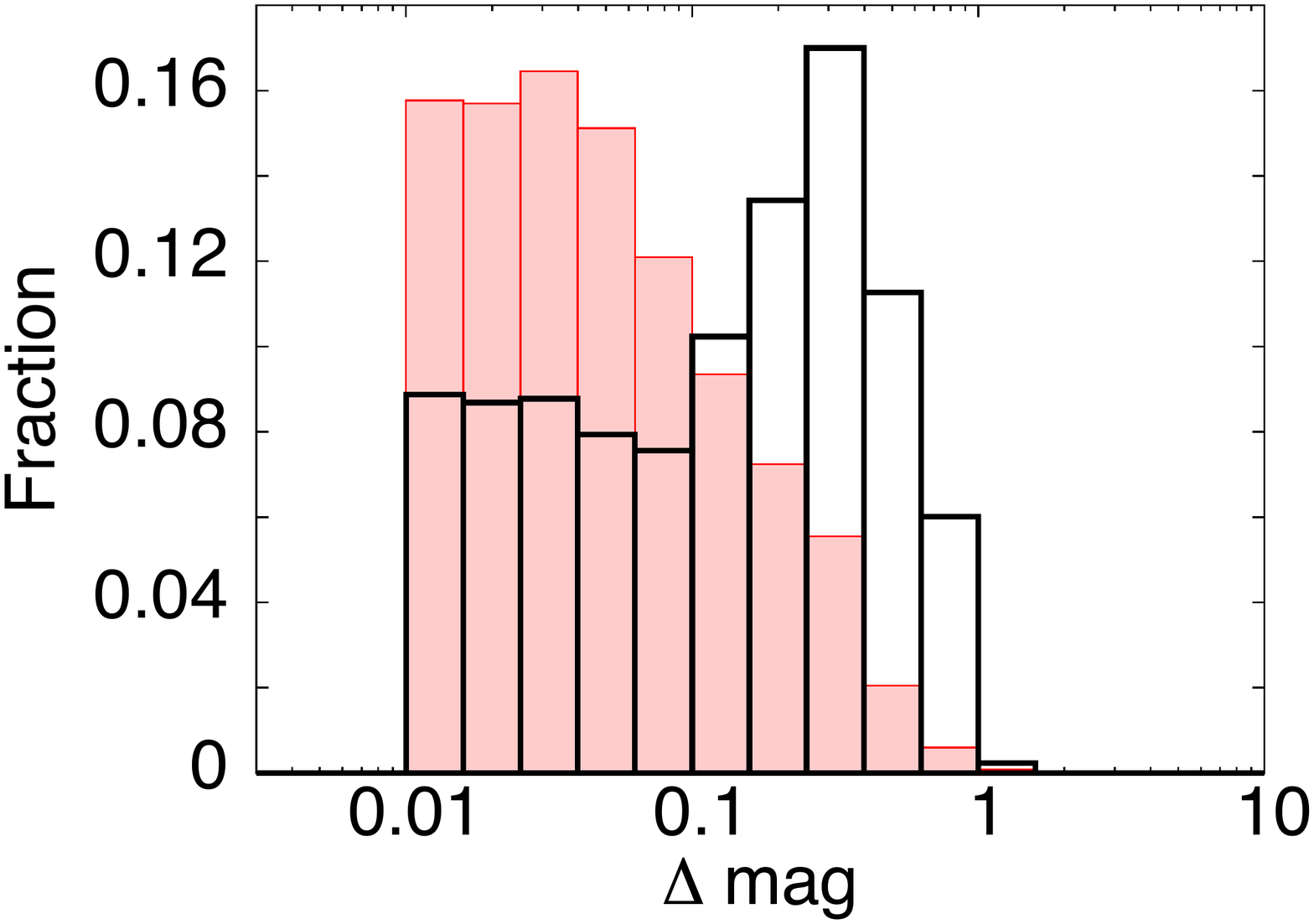}}
\resizebox{2.3in}{!}
	{\includegraphics{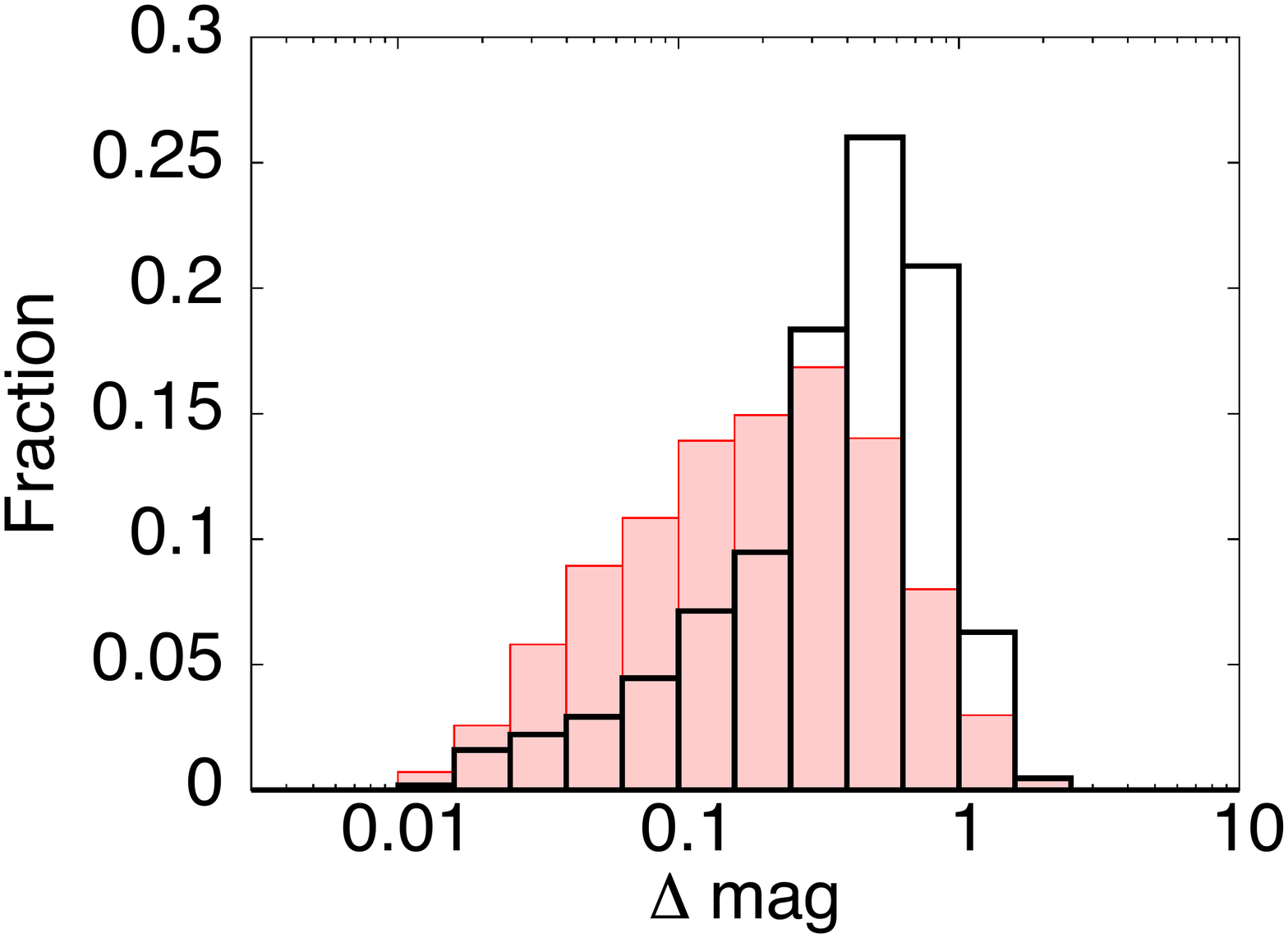}}
\resizebox{2.3in}{!}
	{\includegraphics{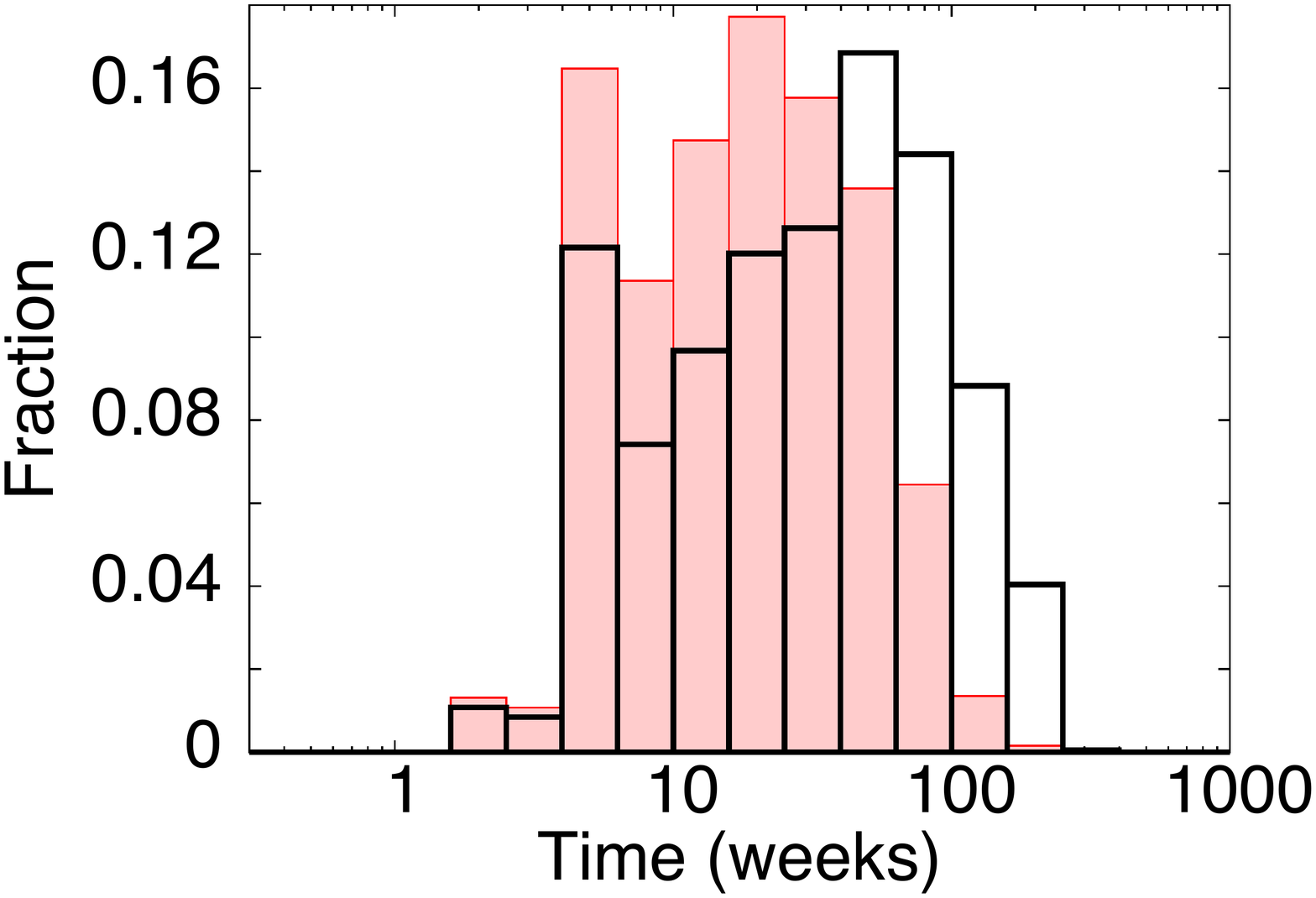}}
\caption{Statistics for `valleys' in saddle point realizations {\it (top)} and minimum point realizations {\it (bottom)}.  Each event consists of a falling side, a minimum, and a rising side.  {\it Left:}  The smaller change in magnitude over the event (either the rising side or the falling side).  {\it Center:}  The larger change in magnitude.  {\it Right:}  The duration of the event in weeks.  The filled red bars represent the realizations with subsolar mass halos, while the black lines represent the realizations without.  \label{fig:lc_valley}}
\end{figure*}

\begin{figure}
\epsscale{1.15}
\centering
\plottwo{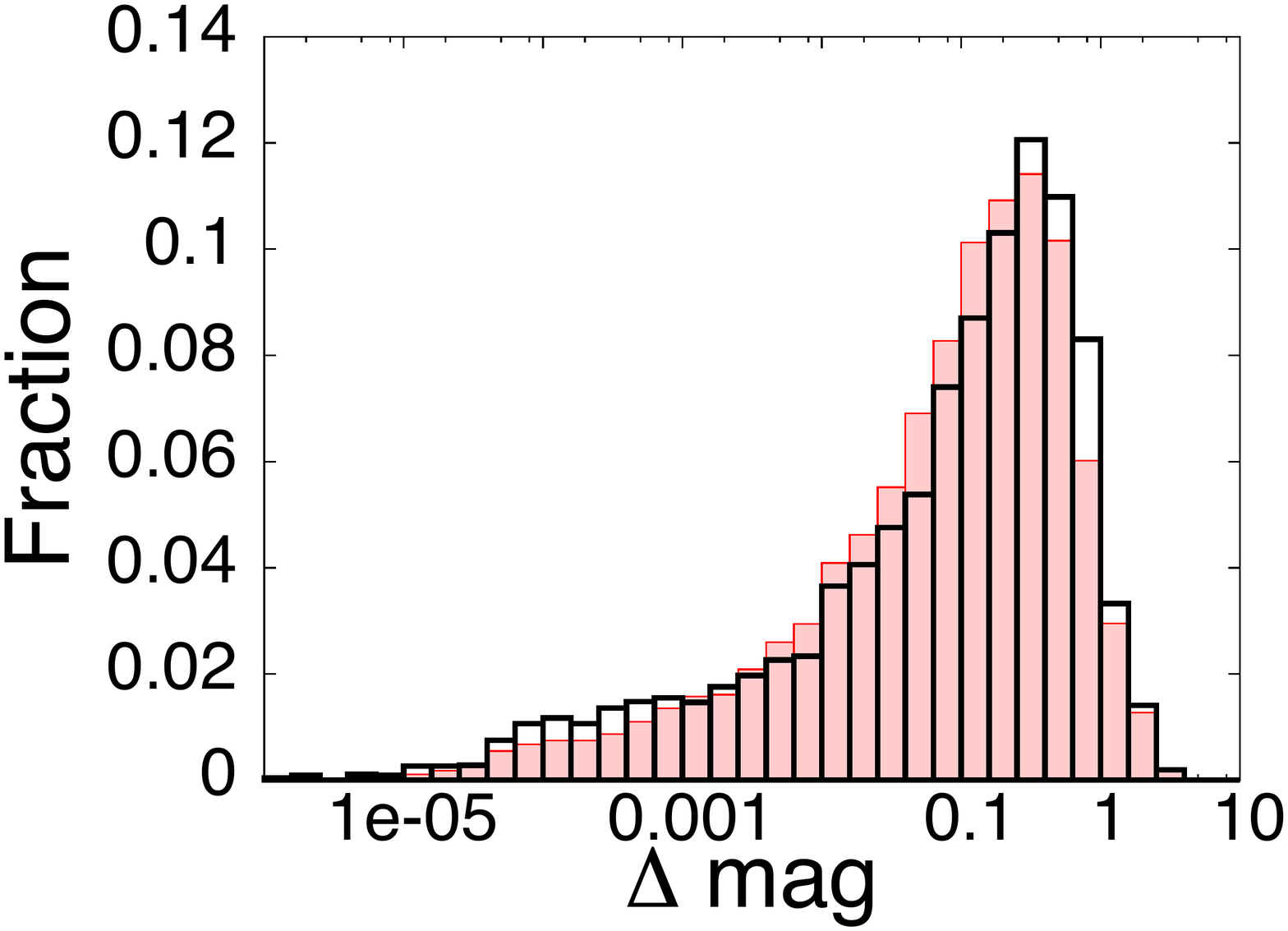}{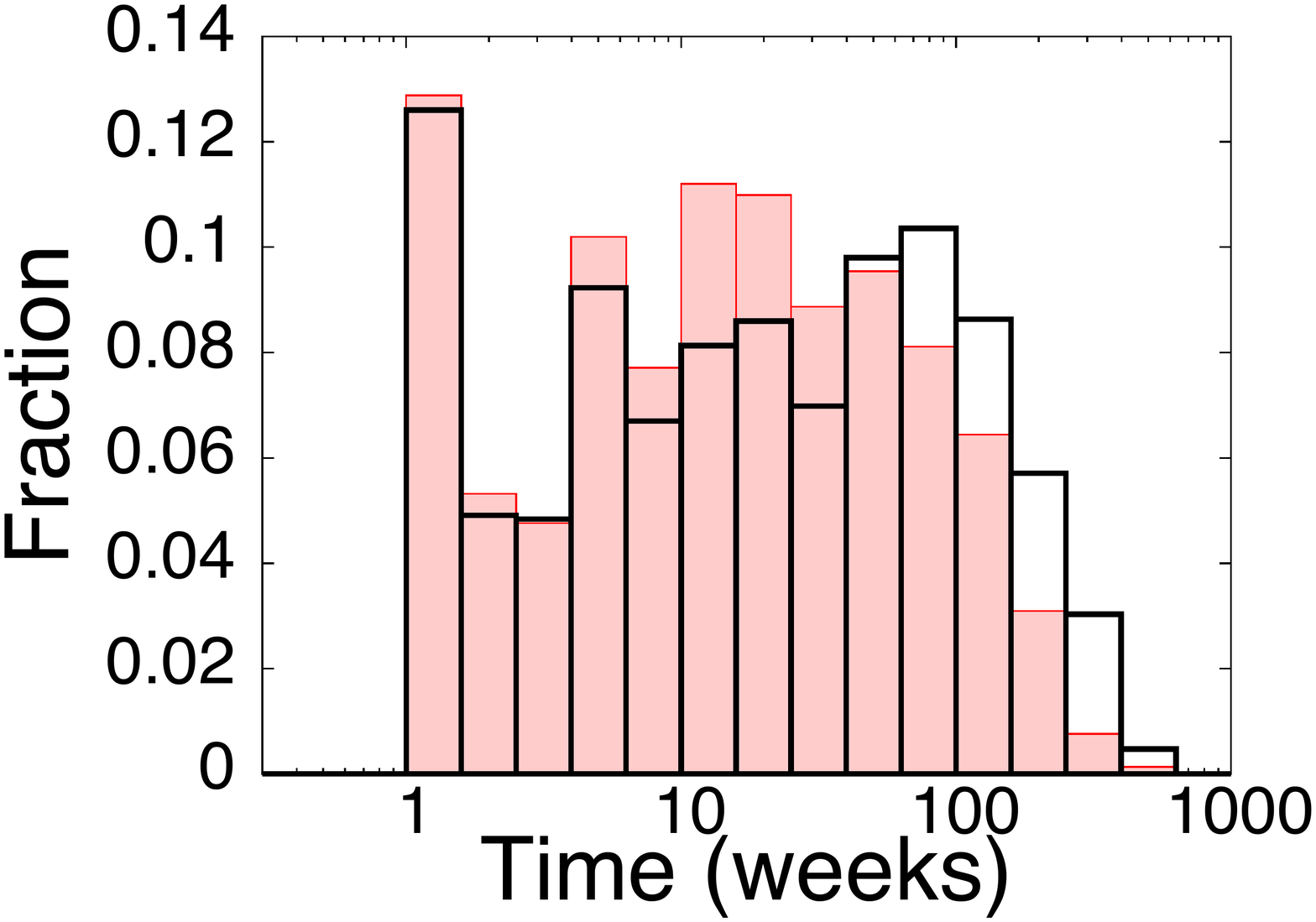} 
\plottwo{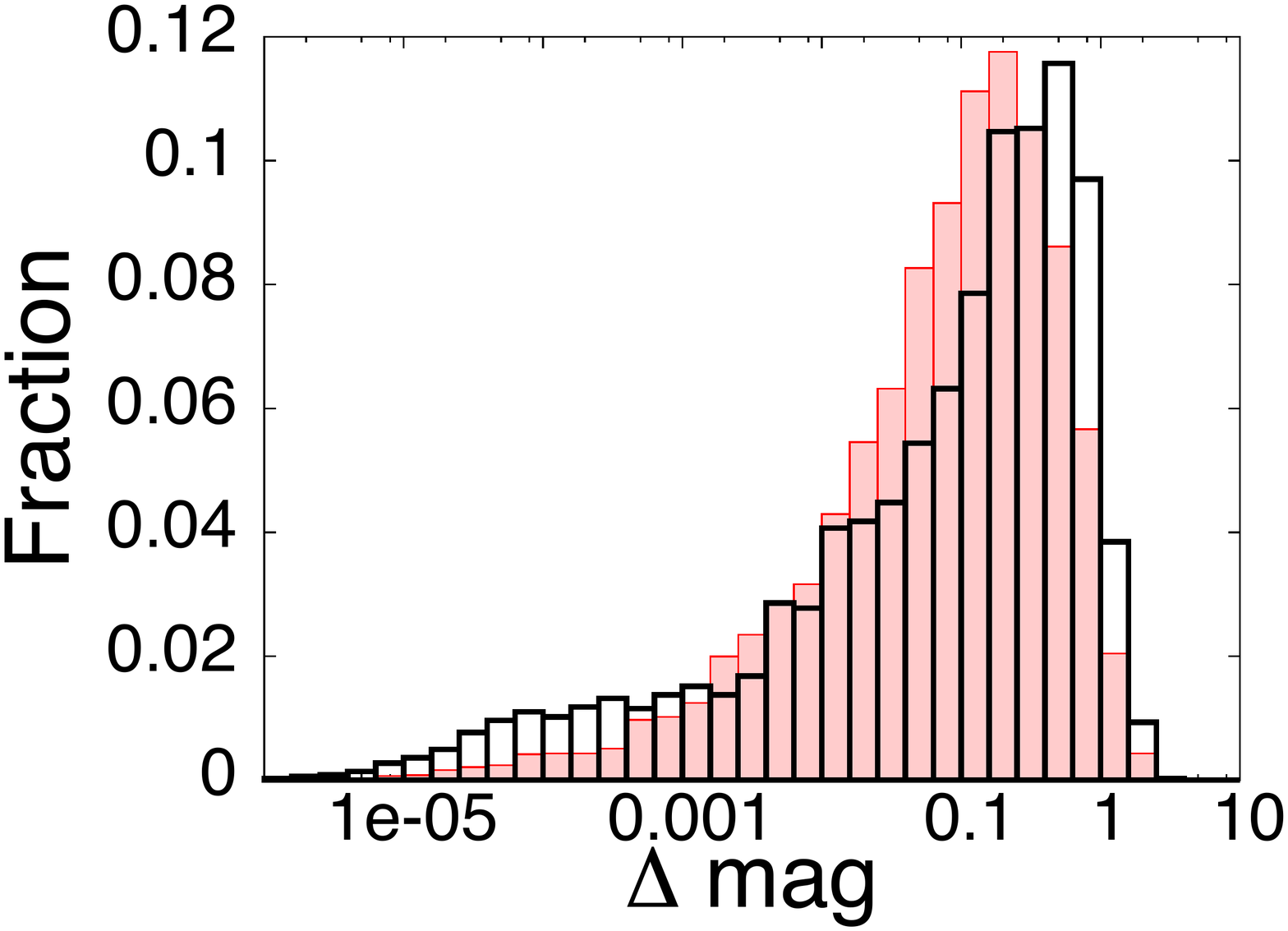}{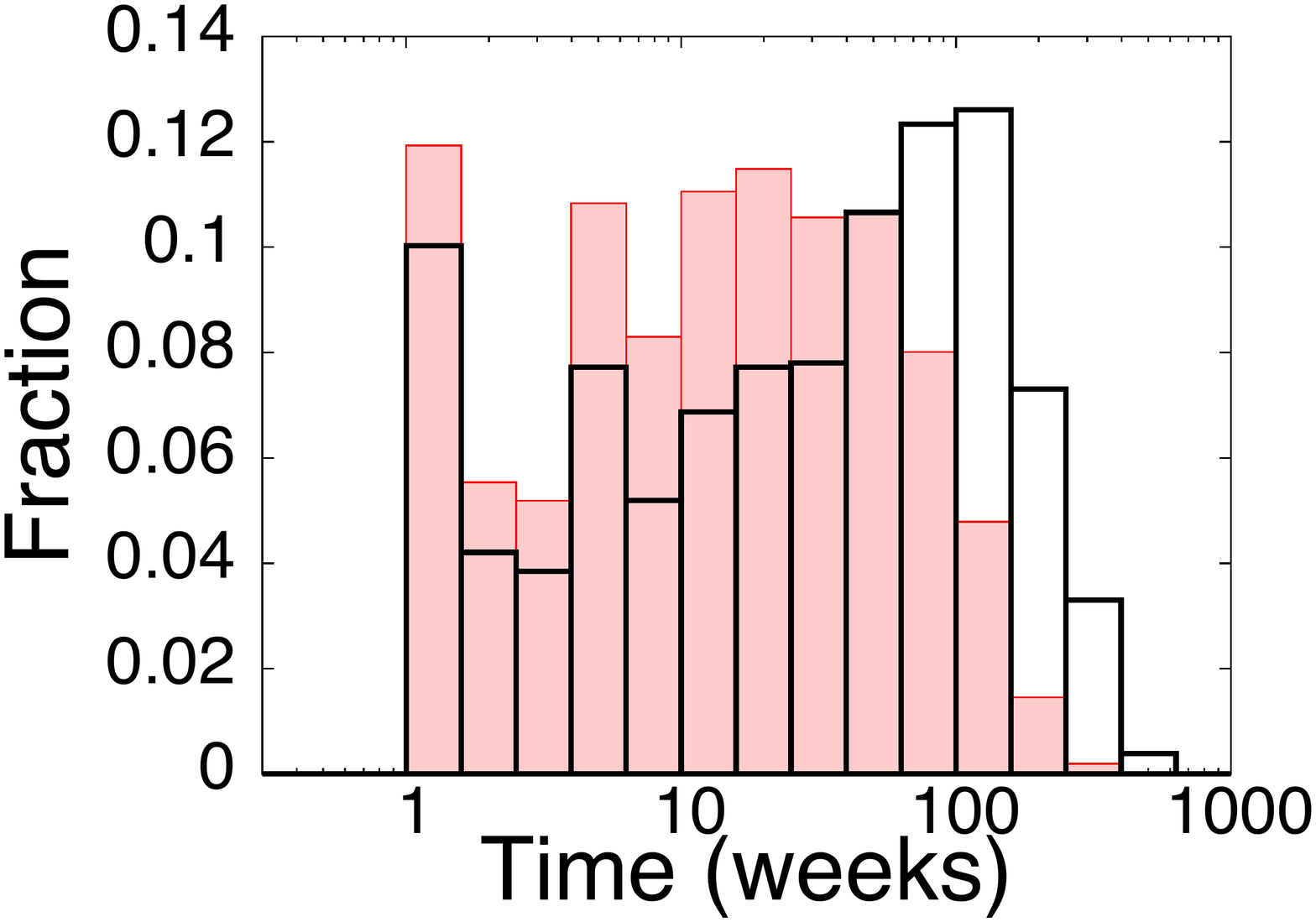} 
\caption{Statistics for `plains' in saddle point realizations {\it (top)} and minimum point realizations {\it (bottom)}.  {\it Left:}  The change in magnitude over the event. {\it Right:}  The duration of the  event in weeks.  The filled red bars represent the realizations with subsolar mass halos, while the black lines represent the realizations without. }
\label{fig:lc_plain} 
\end{figure}

Results for the magnitude amplitude and timescale of events is shown in Figures  \ref{fig:lc_peak}, \ref{fig:lc_valley}, and \ref{fig:lc_plain}.  The peaks show that the magnitude amplitude of the event is smaller for the realizations with dark matter halos than those without, although the timescales are similar.  Similar results are found in the plains.  A much larger difference, however, is seen in the valleys.  Here the minimum magnitude difference is much smaller for the realizations with dark matter halos than for those without.  A milder effect is seen in the maximum magnitude difference and in the timescale;  the magnitude difference and the timescale are somewhat smaller in the case with dark matter halos than in the case without.  Here it seems that when distinguishing between microlensing events and events due to subsolar mass dark matter halos, valley events will be more useful.  

\begin{table}
\begin{center}
\begin{tabular}{c|c|c|c}
Realizations &  Peaks &  Valleys  & Plains  \\
\hline
S1-S7 & 5621 & 4332 & 5774 \\ 
Sh1-Sh7 & 7774 & 6673 & 8445 \\
M1-M5 & 3170 & 2130 & 3639 \\
Mh1-Mh5 & 6027 & 5027 & 6268\\
\end{tabular} 
\caption{Event Statistics}
\label{tab:events}
\end{center}
\end{table}

\begin{figure}
\begin{center}
\includegraphics[height=3.9cm]{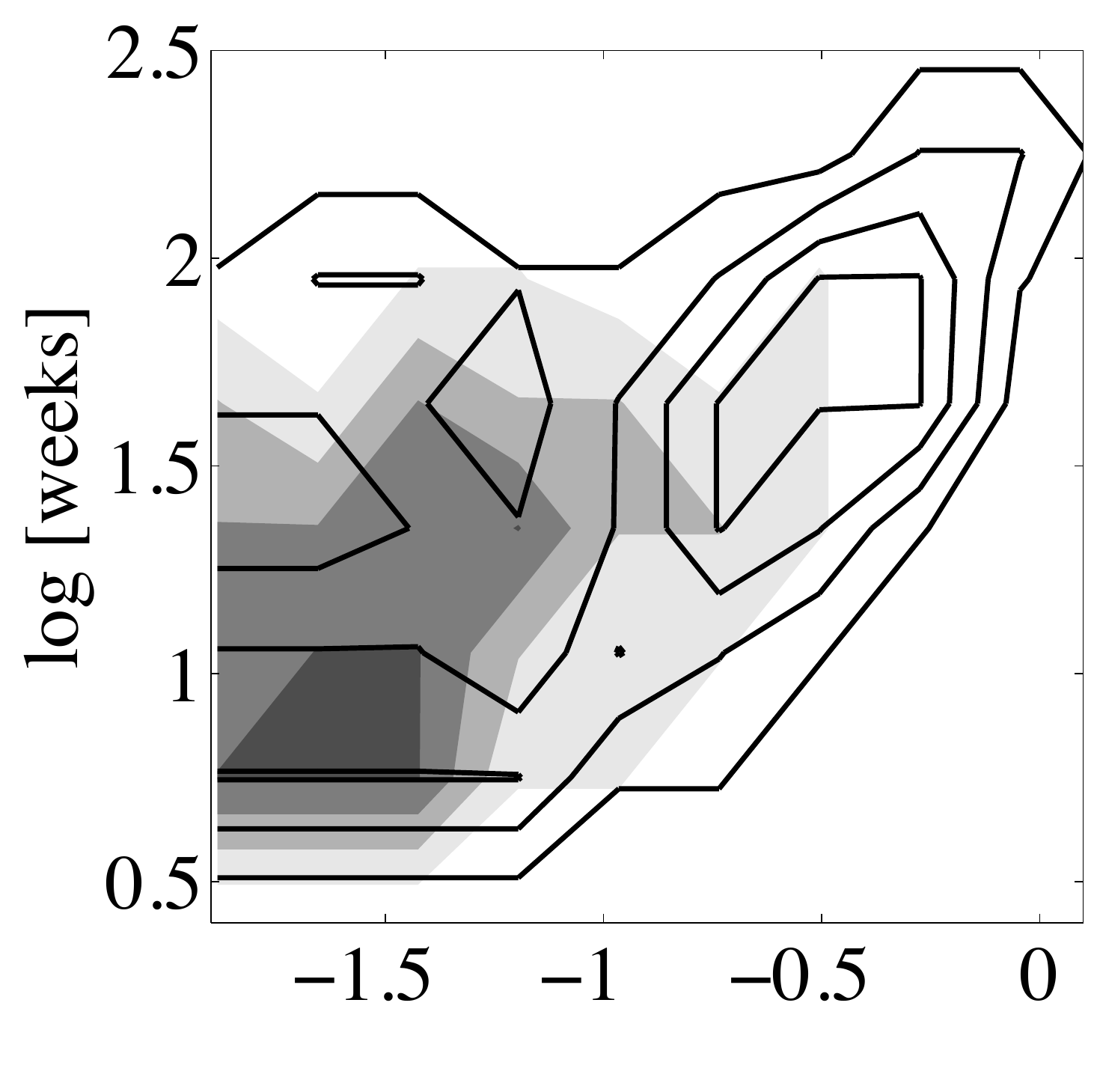}
\includegraphics[height=3.9cm]{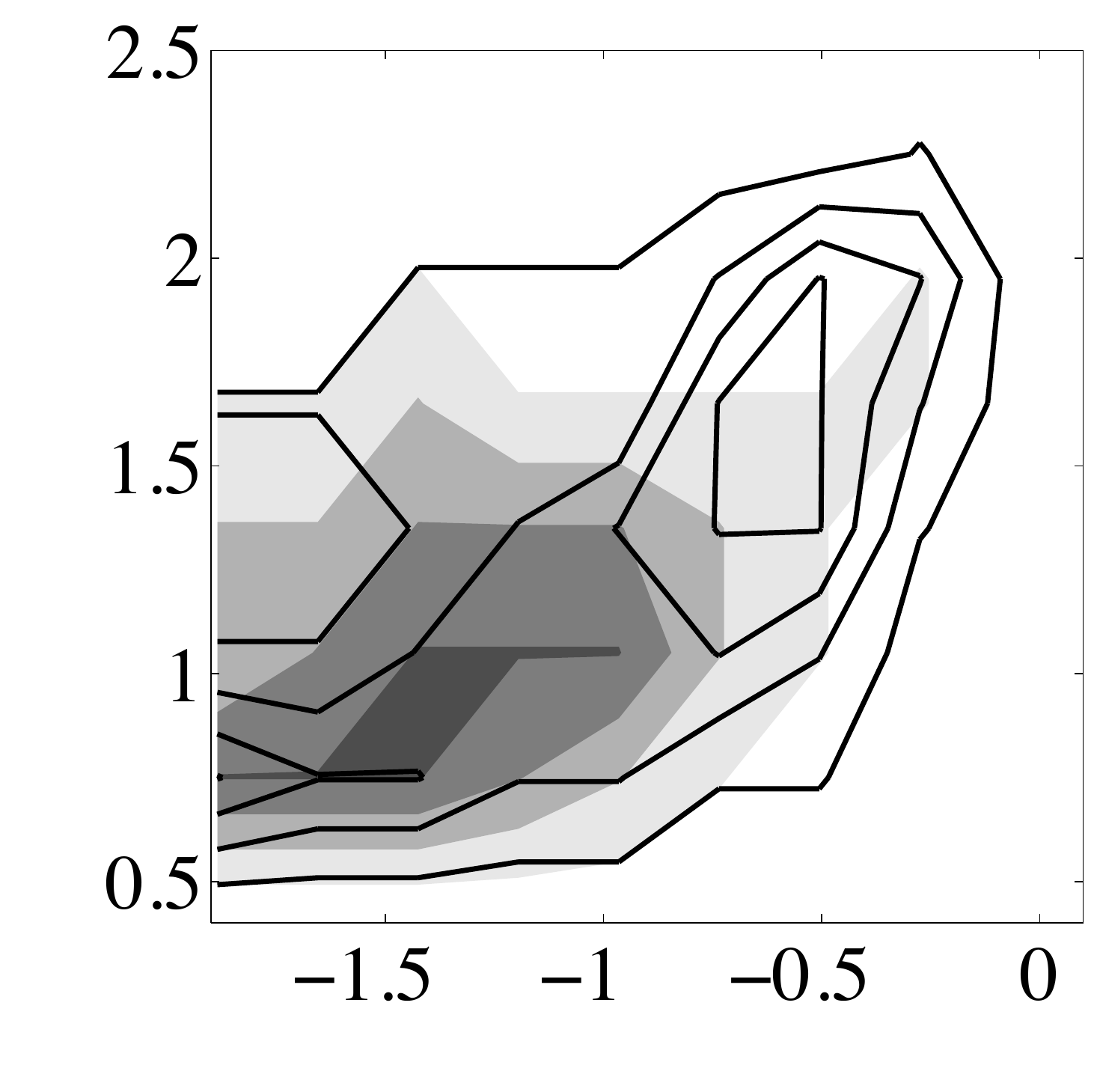} \\
\includegraphics[height=3.9cm]{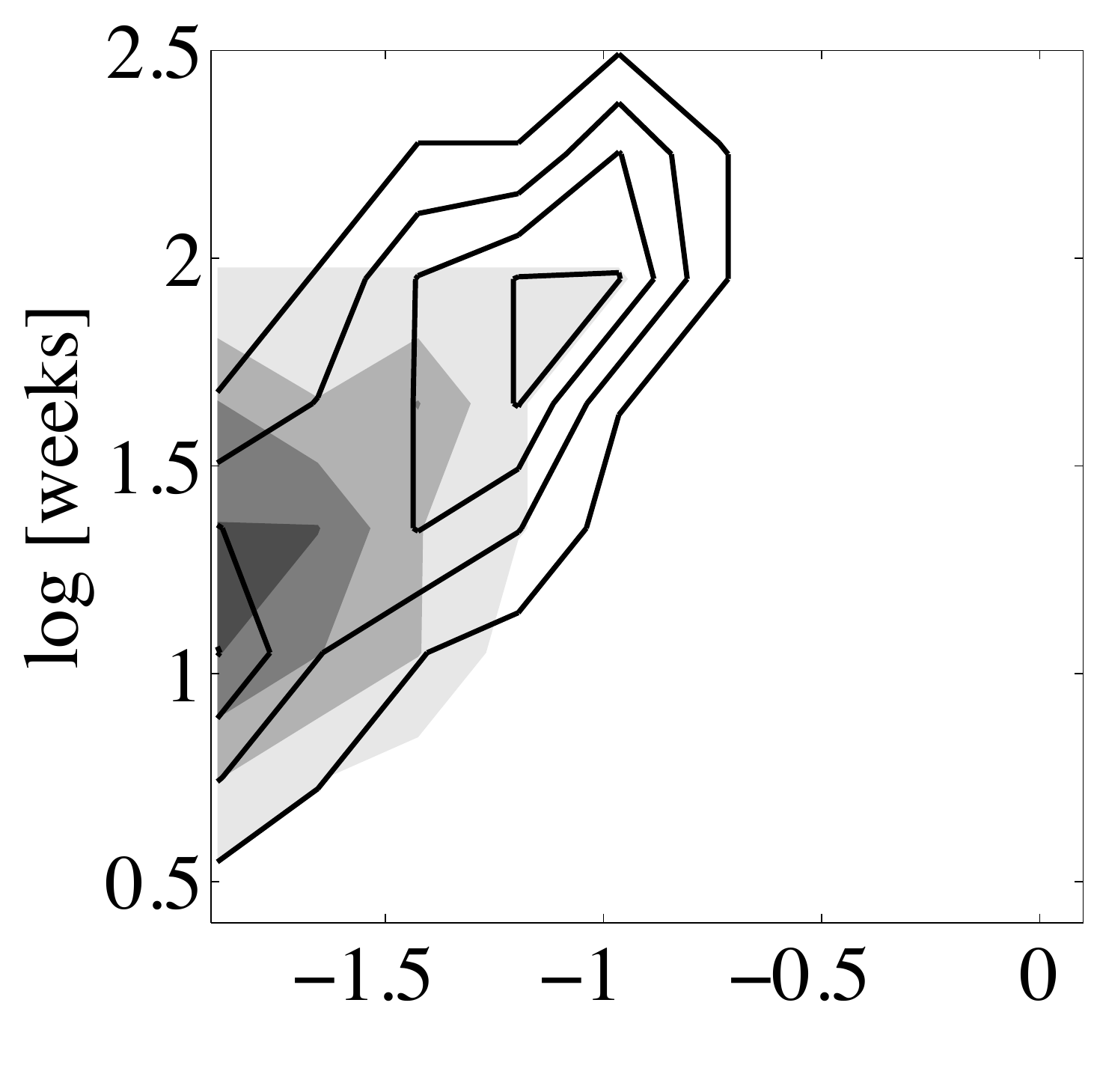}
\includegraphics[height=3.9cm]{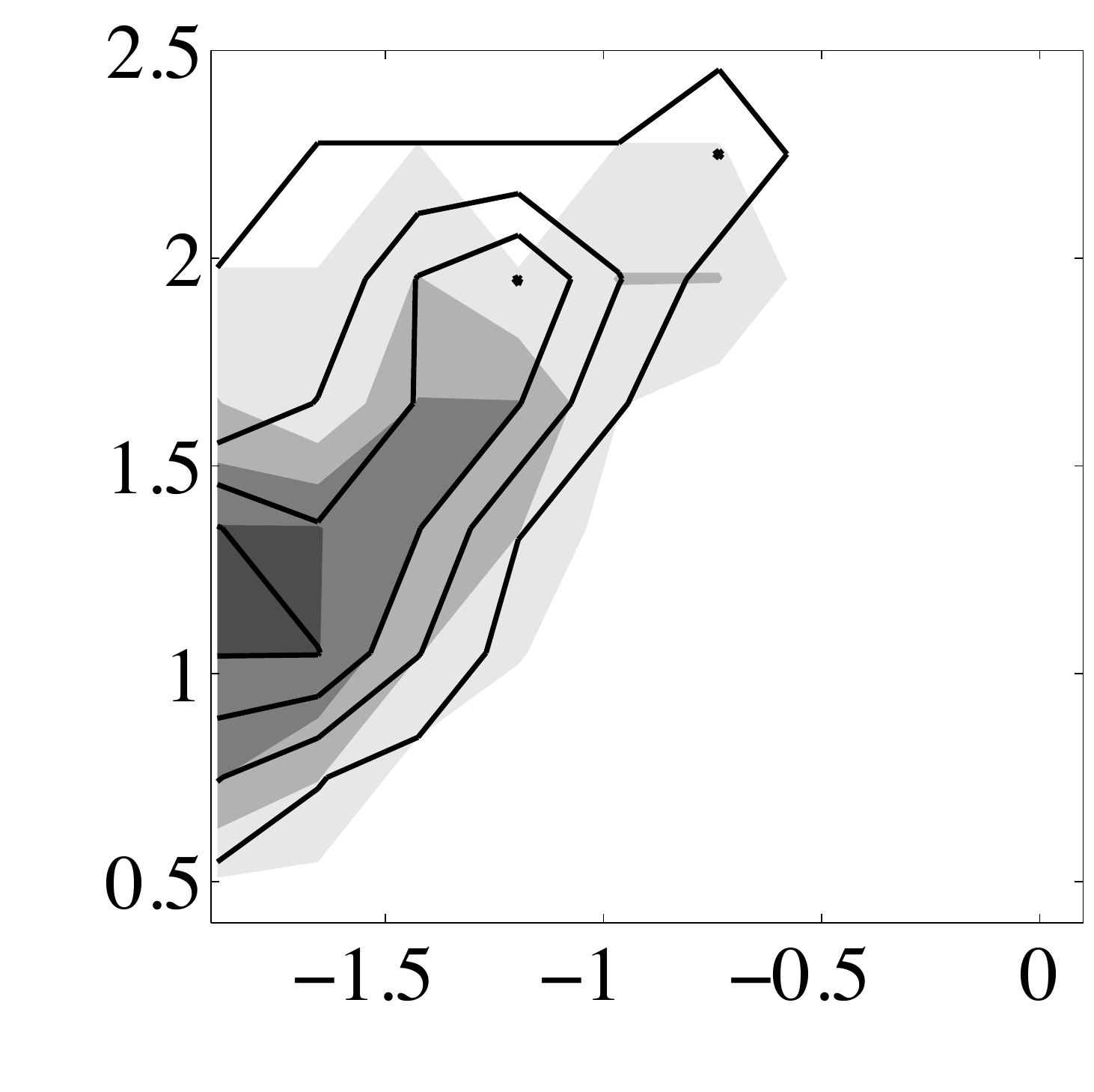}\\ 
\includegraphics[height=3.9cm]{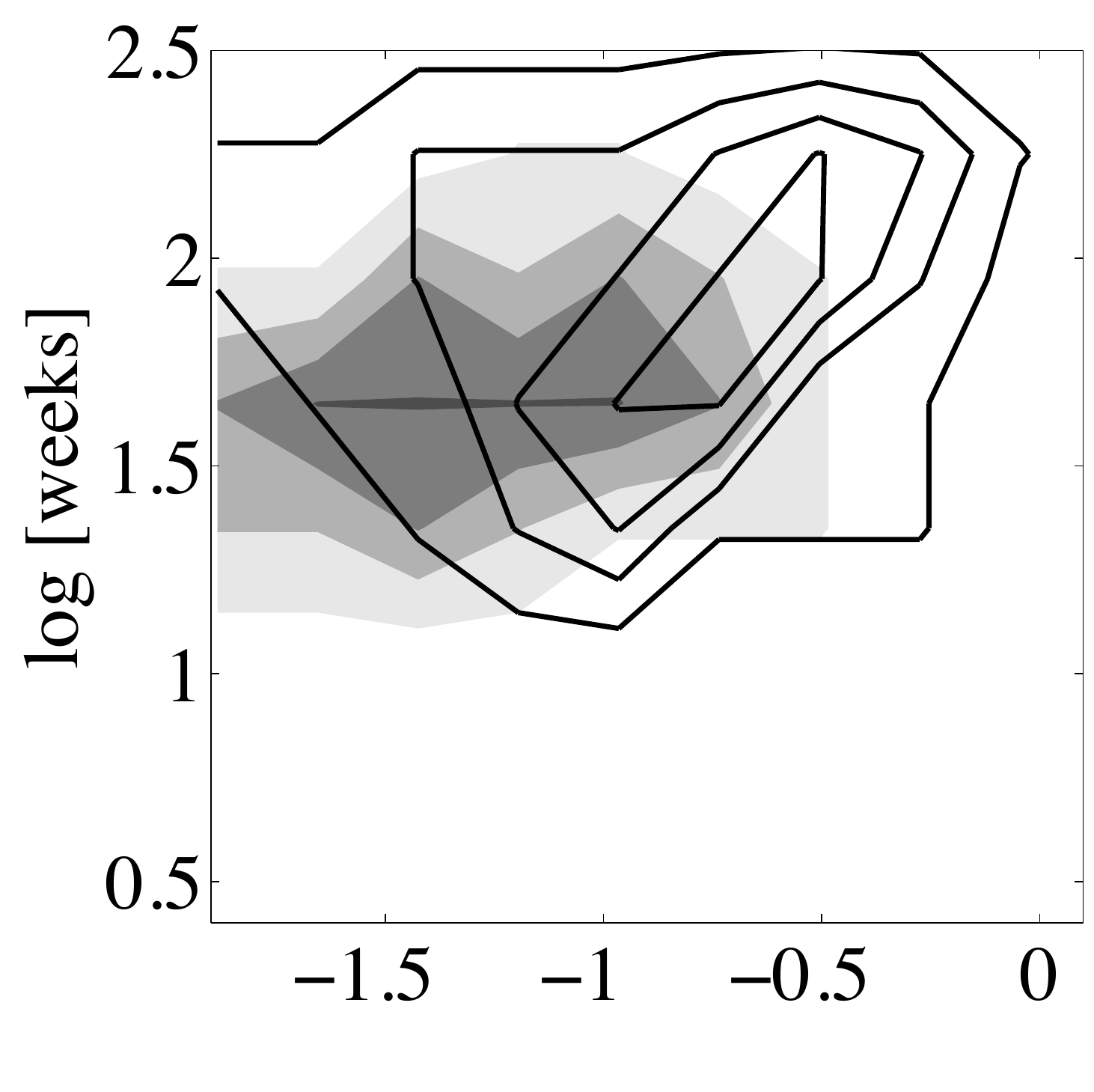}
\includegraphics[height=3.9cm]{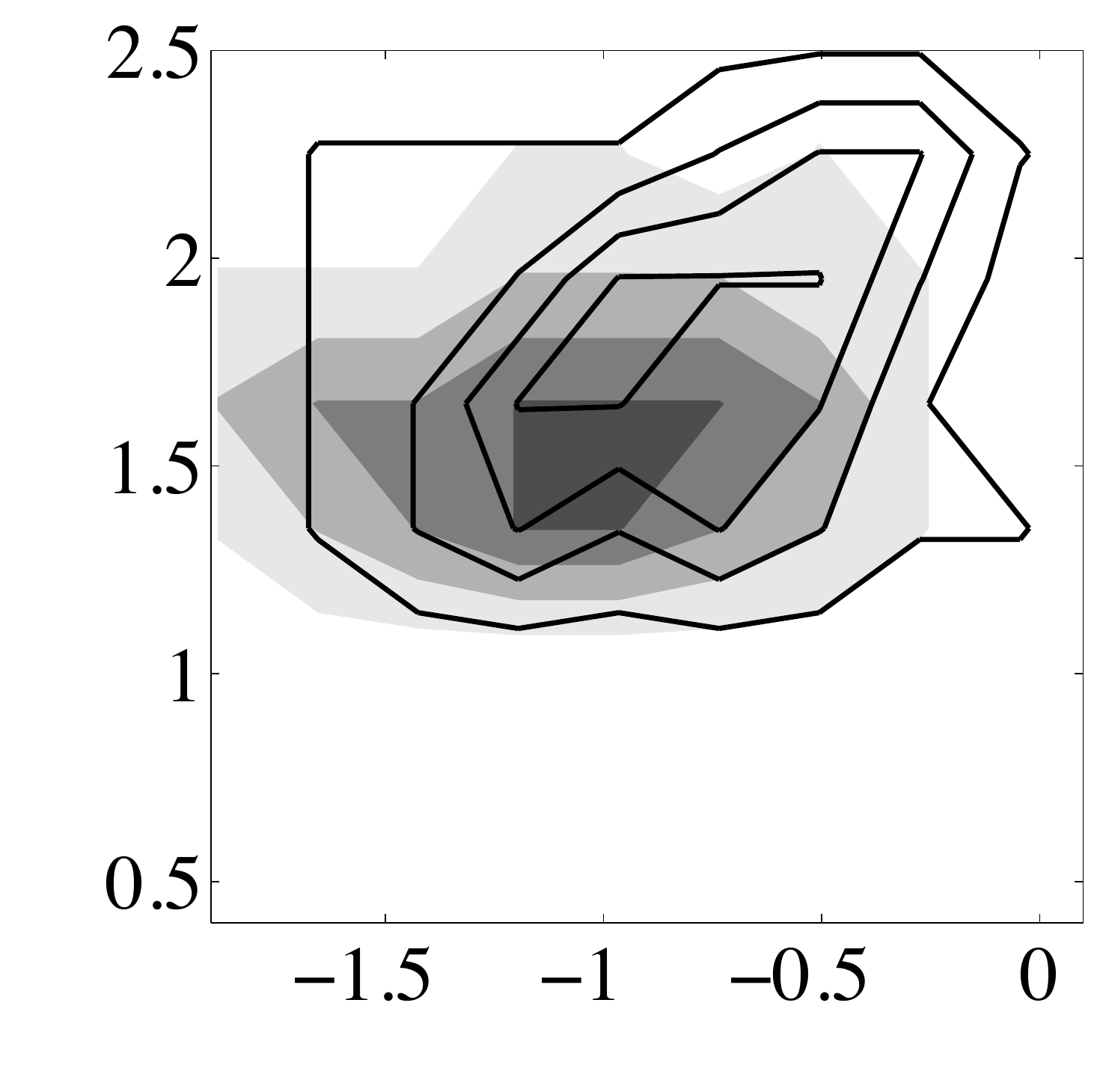} \\
\includegraphics[height=3.9cm]{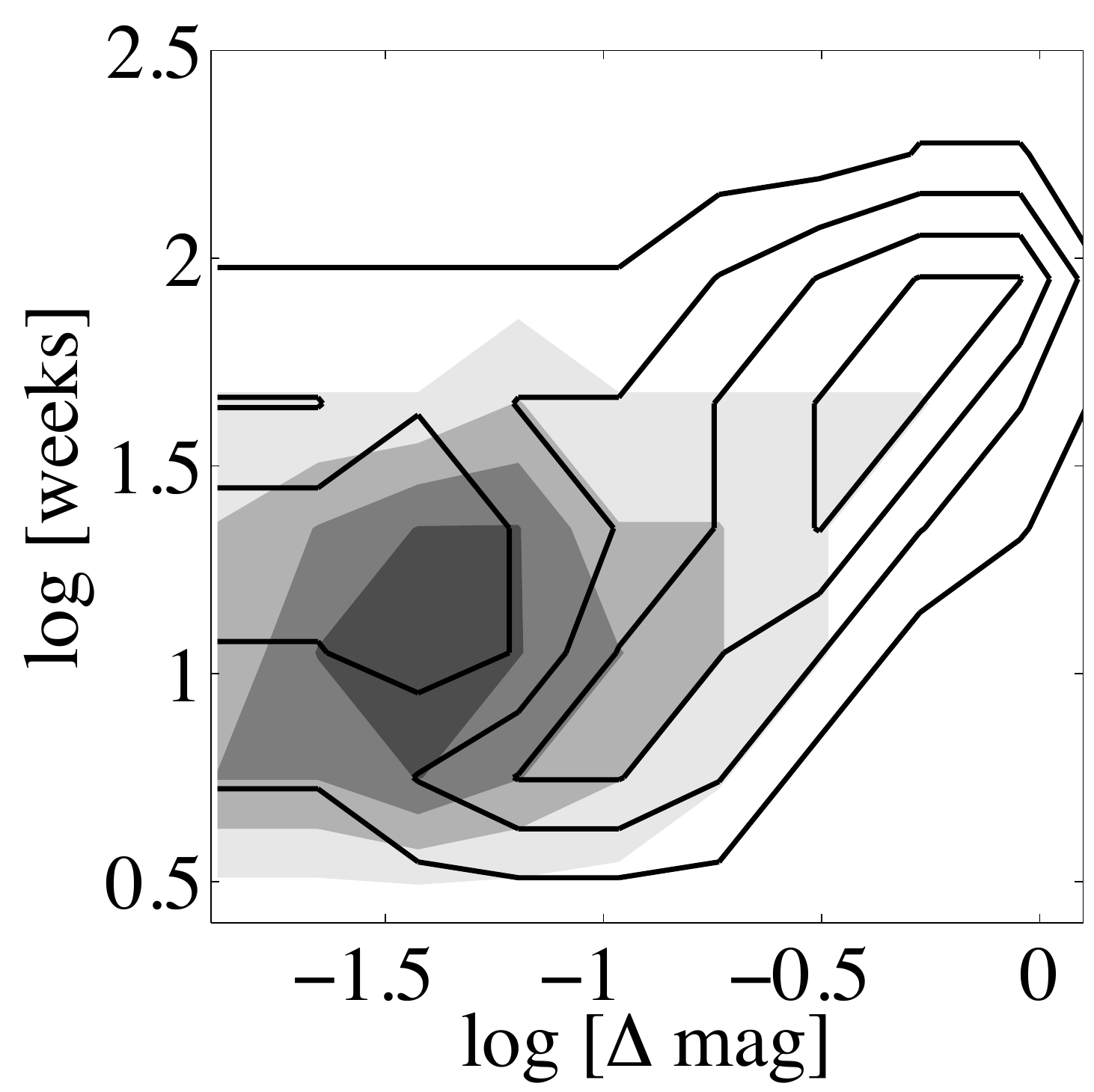}
\includegraphics[height=3.9cm]{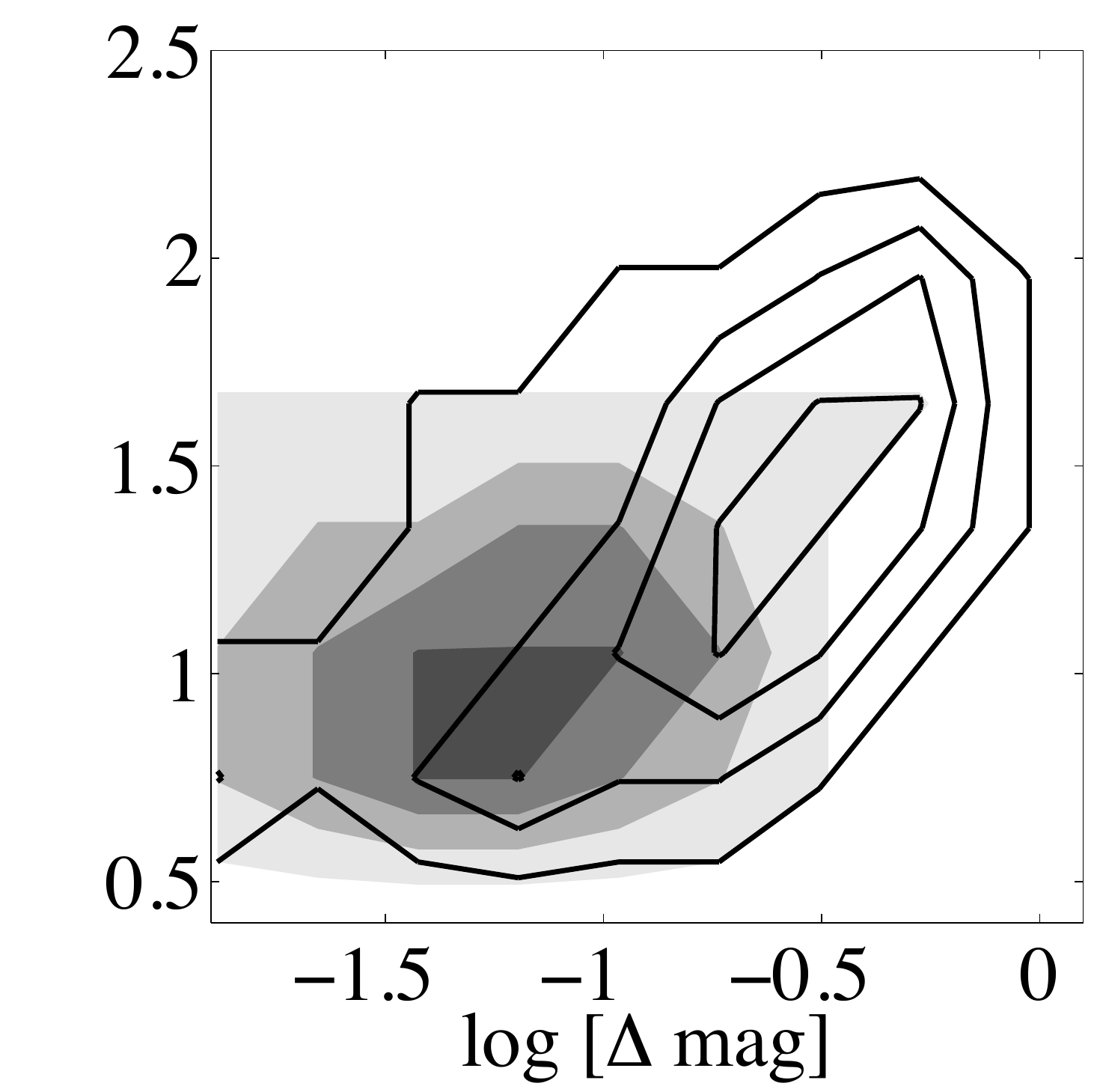} 
\caption{The minimum magnitude amplitude - timescale plane for minimum source points {\it (left)} and for saddle point {\it (right)}.  The realizations with dark matter halos are shown in greyscale and the realizations without are shown in contour lines.  In both cases, the contours show where 90\%, 70\% , 50\%, and 30\% of the events lie.  From the top, the rows show the fiducial case, the case for which the source radius is 8 times larger, the case for which the observation frequency is 6 weeks, and the case for which the source velocity is 900 km/s.  }
\label{fig:contours} 
\end{center}
\end{figure}

These differences can be summarized in the top panels of Figure \ref{fig:contours}.  Here the minimum magnitude amplitude and the timescale for valleys are plotted in contours.  The realizations with and without dark matter halos are found in overlapping regions.  This is unsurprising, since a large fraction of the the events in any realization is due to stars.  As we consider smaller fractions of the events, we see that 50\% of the events are found in different parts of the parameter space.  Events due to dark matter halos are found with minimum magnitude amplitudes less than 0.1 mag and event timescales less than a few months.  Events due solely to stars display a range of minimum magnitude differences and timescales.  This suggests that it would be impossible to classify a single event as being produced by dark matter halos or by stars.  At the 50\% level, though, valleys in realizations without dark matter halos have a minimum magnitude difference $\sim 0.3$ and timescales on order of years.  

Figure \ref{fig:contours} also shows the results of varying some of the fiducial conditions, such as source size, observation frequency, and source velocity.  As the source size increases, events are smoothed out.  Fewer total events will be measured, and the remaining events have smaller magnitude amplitudes.  A distinct signal from dark matter halo events will be more difficult to discern.  A less frequent monitoring of a lensing system will also obscure the differences between stars and dark matter halos;  observations will be sensitive to only longer duration events.  A faster effective source velocity (900 km/s) may have similar effects to less frequent observations -- although, in this case, the effect is mild --  but it will sample from a larger area in the source plane (and more events) than the fiducial velocity can.

\begin{figure*}
\begin{center}
\plottwo{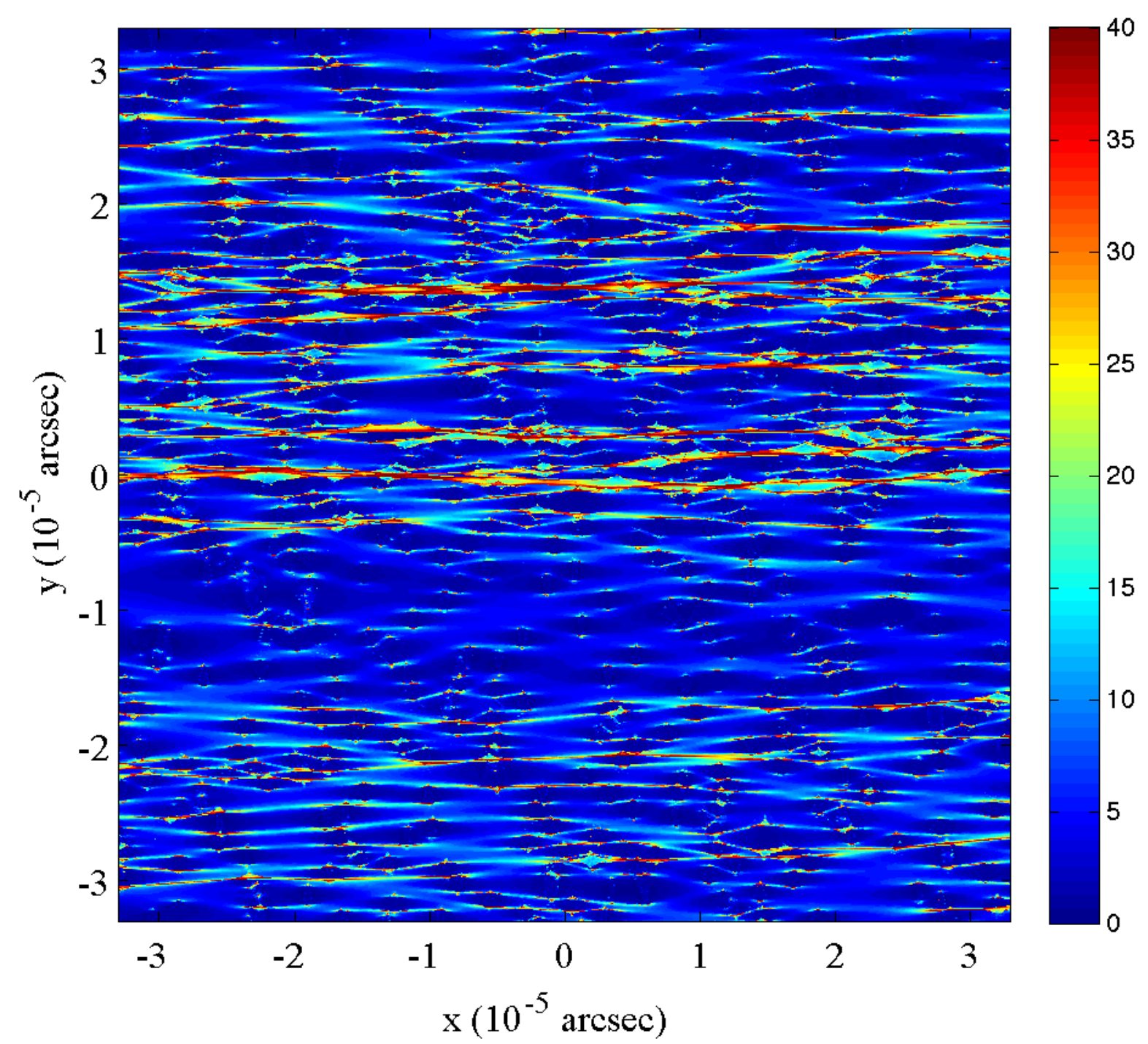}{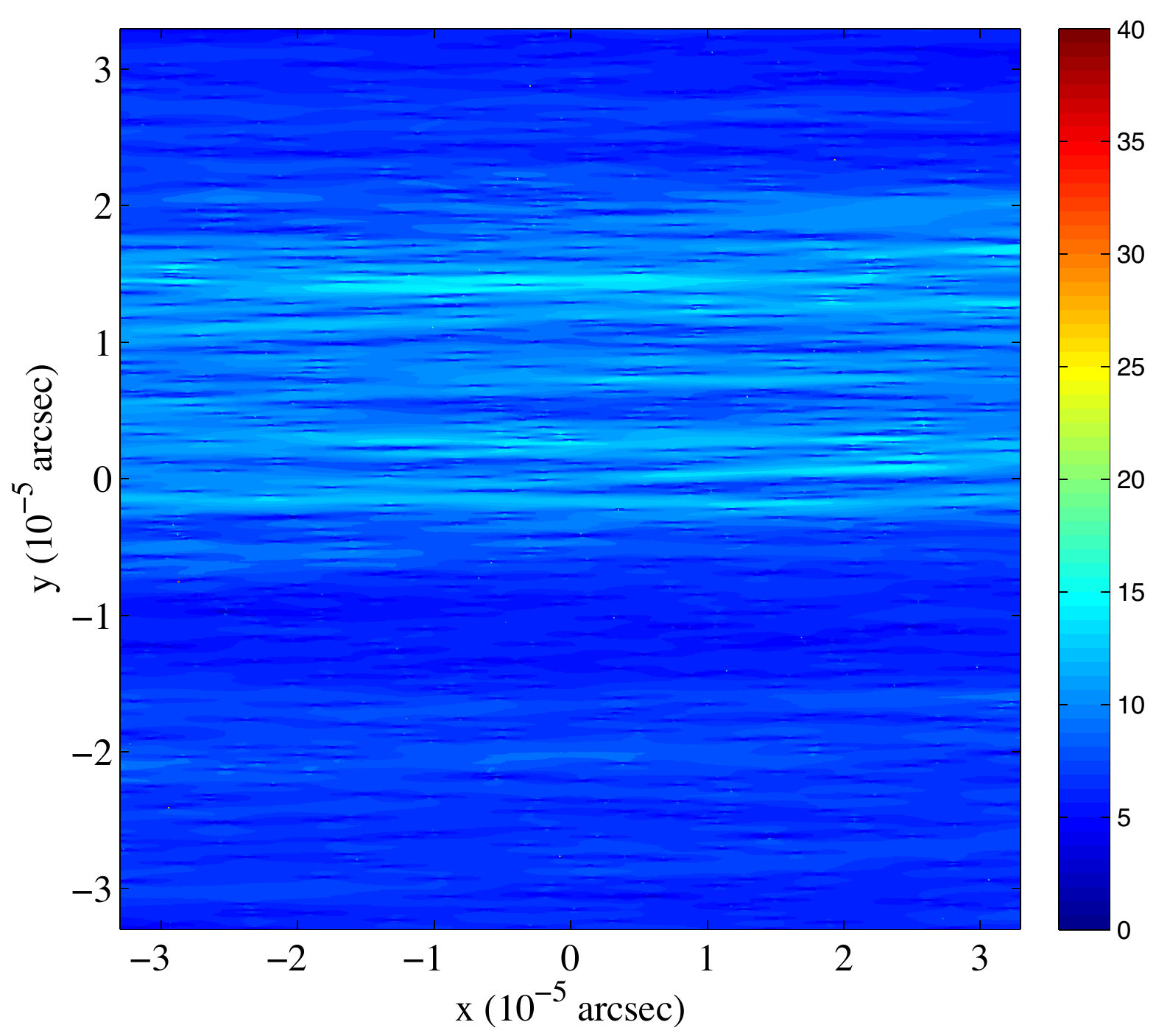}
\caption{Sample magnificaton maps for a saddle point image in which all the surface mass density is in subsolar mass dark matter halos with power-law surface density profiles of $\Sigma \propto R^{-1.5}$ {\it (left)} and $\Sigma \propto R^{-1}$ {\it (right)}.  \label{fig:magmap_profile}}
\end{center}
\end{figure*}

\section{Discussion}
\label{sec:discussion}

The light curve analysis shows that subsolar mass dark matter halos in galaxies leave clearly observable effects.  They create `nanolensing' events that are seen as small magnitude amplitude and short duration events in a light curve.  The amplitude in magnitude is a less than $\Delta$ mag $=0.1$ and require observations with small photometric errors.  The timescales are significantly less than 1 year and require a short cadence of observations.  

There remains some challenges that may impair our ability to detect subsolar mass dark matter halos with gravitational lensing such as the dark matter halo profile, the fraction of surface density $\kappa$ in stars and in subsolar mass halos, and the intrinsic variability of quasar sources.

As discussed previously, our fiducial model uses  $\Sigma \propto R^{-1.5}$ which is steeper than the $\rho \propto r^{-1.5}$ that most recent simulations suggest, although the possibility still remains that these dark matter halos will have higher central densities than calculated \citep{ishiyama_etal10}.  The strong `nanolensing' effects of subsolar mass dark matter halos are dependent upon the density profile.  Figure \ref{fig:magmap_profile} shows magnification maps in which the entire surface mass density is put into subsolar mass dark matter halos using the fiducial mass function.  The panel on the left uses the fiducial $\Sigma \propto R^{-1.5}$, while the one on the right uses $\Sigma \propto R^{-1}$.  These maps are lower resolution and show a larger area than in the realizations used in the light curve analysis, but it is clear that a shallower density profile will result in smaller structures, which will be more difficult to detect.  

While the density profile of subsolar mass halos has a significant effect on the size of features at fixed mass function and surface mass density, the same kinds of features in magnification maps -- peaks, valleys, plains -- appear regardless of the profile.  In addition, any given magnification map is not a unique representation of the profile of the subsolar mass halos.  For example, the two panels in Figure \ref{fig:magmap_profile} could be made similar by increasing the total mass of the subsolar mass halos in right panel.  This degeneracy is not confined to extended halos either.  The magnification map created by a distribution of fiducial subsolar mass halos can be approximated by a distribution of subsolar mass point-like objects with a smaller surface mass density.  

The fraction of surface density in stars and in dark matter halos will also impact the observability of subsolar mass dark matter halos.  As the fraction of surface density in stars increases, there will be a smaller amount of mass that can be attributed to subsolar mass dark matter halos.  
If the stellar fraction decreases, more mass can be attributed to dark matter halos.  The microlensing fluctuations, however, will remain significant until the stellar fraction drops below $\sim$10\%.  

Ignoring the effect of stars, as we vary the surface mass density in dark matter between a smooth component and a component of dark matter halos, we will see similar effects as described by \citet{schechter_wambsganss02}, with only the size of the events on the source plane scaled down.  As the fraction in dark matter halos grows, a growing number peak events interlace, crowding out the deep valleys seen in the saddle point magnification maps.  Given that subsolar mass halos are not point sources, this effect will never been as strong as in the case described by \citet{schechter_wambsganss02}.  As the fraction of dark matter halos decreases toward 0\%, the peaks become less frequent, more distinct, and separated by large plains or valleys.  

An additional source of variability in observed light curves is the intrinsic variation of the quasar source.  In observed lenses, light curves from different images of the same source are generally differenced in order to eliminate the effect of intrinsic variability.  However, this obscures the differences between peak and valley events due to lensing.  We can compare the variability due to lensing to the intrinsic variability by studying a sample of unlensed quasars.  \citet{schmidt_etal10}  characterize a sample of quasars taken from the Sloan Digital Sky Survey (SDSS) Stripe 82.  For each object, they quantify the mean variability amplitude as a function of the time between observations by calculating the structure function \citep{1996A&A...306..395C,1999MNRAS.306..637G},
\begin{equation}
V(\Delta t) = \left\langle\sqrt{\frac{\pi}{2}} | \Delta m_{i,j} | - \sqrt{\sigma^2_i + \sigma^2_j}\right\rangle_{\Delta t},
\label{eq:sf}
\end{equation}
where the average $\langle \rangle_{\Delta t}$ is taken over all pairs of observations $i$ and $j$, whose time difference falls into the bin $\Delta t$.  \citet{schmidt_etal10} model the shape of the structure function for a sample of quasars as a power law with amplitude $A$ and slope $\gamma$, $V_{\rm model} = A \left( \Delta t /{ \rm 1 year}\right)^{\gamma}$ and find that $0.07 < A < 0.25$ and $0.15 < \gamma < 0.5$ contain over 90\% of the quasars.   

\begin{figure}[t]
\resizebox{3.5in}{!}
	{\includegraphics{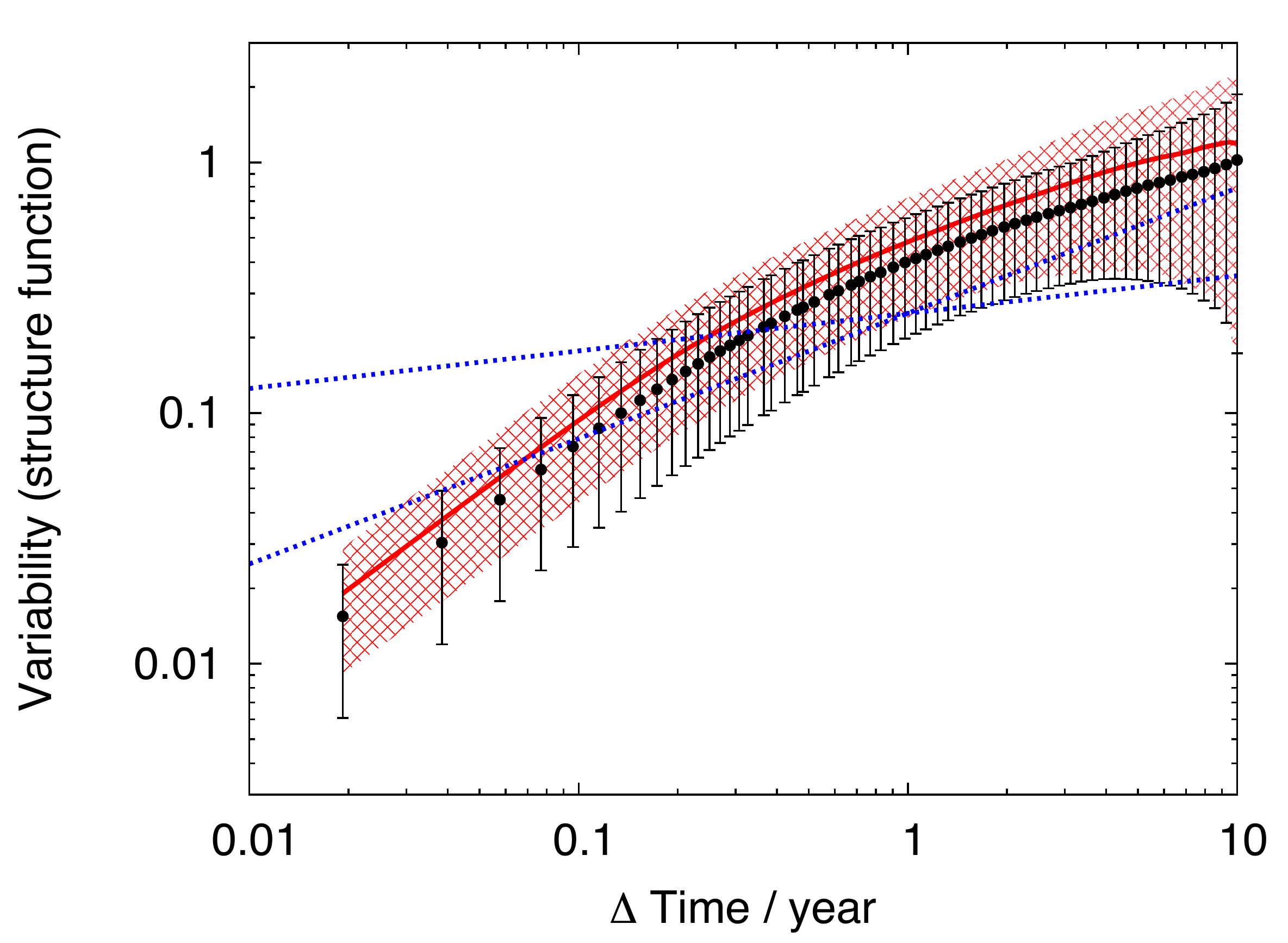}}
\caption{The mean and the standard deviation of the structure function for mock light curves with subsolar mass dark matter halos (red) and without (black) calculated by Equation \ref{eq:sf}.  The blue lines show power laws with amplitude, $A=0.25$ and slope, $\gamma = 0.15$ and $0.5$ (see text).  \label{fig:struct_fn}}
\end{figure}

We calculate the structure function for light curves with and without subsolar mass dark matter halos.  The results for the mean and the standard deviation are shown in Figure \ref{fig:struct_fn} with comparison lines for the power laws with the largest amplitudes, $A=0.25$ and the bounds of the slopes $\gamma = 0.15$ and $\gamma=0.5$.  It is clear that the structure function is not particularly sensitive to the differences between lensing due to stars and lensing due to subsolar mass dark matter halos.  We can see, though, that lensing causes a larger amplitude in variability compared to intrinsic variability at scales of $\gtrsim 0.3$ years.  At the smallest scales of $\lesssim 0.1$ years, on the other hand, the signal from intrinsic variability will dwarf that of lensing and image differencing will be necessary.  At at intermediate scales -- where we may hope to find nanolensing events --  it is unclear if intrinsic variability could be confused with lensing events.   

We create a sample of color-selected quasars and a comparison sample of color-selected F/G stars using criteria based on those in \citet{schmidt_etal10} using the SDSS \citep{abazajian_etal09,york_etal00,gunn_etal06,gunn_etal98,fukugita_etal96,hogg_etal01}.  The cadence of observations of Stripe 82 is such that each object is sampled  for 2-3 months per year (the median time is $\sim$10 weeks).  This sampling may be long enough to search the quasar light curves for peaks and valleys in a similar manner to that employed on the mock lensed light curves.  Unlike mock observations, however, the $r$-band observations have typical magnitude errors of 0.03 and the $g$-band of 0.04.  We create a simple peak and valley finder that looks for extrema in the $r$-band with durations longer than 1 observation and amplitudes above the observed photometric errors and that have a corresponding event in the $g$-band data.  The quasar sample finds events with median duration of $\sim$5 weeks, while the star sample has a median duration of $\sim$3 weeks.   The duration of quasar events is typically less than the total amount of time sampled, less than the typical duration of micro- and nanolensing events, and greater than what we may presume is noise found in the non-varying star sample.  This could be an indication that events created by intrinsic variability have different properties than those caused by lensing.  Without a more richly sampled dataset with smaller photometric errors, however, any definitive conclusions about the nature of intrinsic quasar variability is not possible.

\section{Conclusions}
\label{sec:conclusions}

Subsolar mass dark matter halos are a prediction of cold dark matter cosmologies and a test thereof.  In addition, measuring the small scale structure of dark matter will aid in understanding structure formation and in making predictions for current and future dark matter particle detection experiments.  

Here we suggest that strong gravitational lensing could be used to detect subsolar mass dark matter halos and show that such halos  create observable lensing effects.  We describe how the effect of subsolar mass dark matter halos differ from the effect of stars in the galaxy lens:
\begin{enumerate}
\item Subsolar mass dark matter halos add lensing events to light curves that are already populated by events from stars.

\item Dark matter halos cause events with preferentially smaller amplitudes and timescales, particularly in the case of `valley,' or dimming, events.  In such events, the minimum magnitude amplitudes are typically less than 0.1 mag and the timescales less than $\sim$100 days.
\end{enumerate}
These effects are visible with a simple analysis of light curves and without knowledge about the larger-scale configuration of the lens that would require detailed lens modeling, such as the shape and size of the galaxy lens in which the subhalos are embedded or the intrinsic flux of the source.

Several observed lens systems have shown variability at timescales less than $\sim$100 days.  These events are often identified as `brightening' events in differential light curves:  light curves from two images are offset by the measured time delay and differenced to eliminate the effect of intrinsic source variability.  Thus, a brightening event attributed to one image of a pair can be interpreted as a dimming event in the other image.  And a single event of $\sim$100 days could be a peak event, a valley event, or some linear combination of both, and it could be due to stars or due to subsolar mass dark matter halos.  

Monitoring of Q2237+0305 showed a single $\gtrsim$100 day event in \citet{schmidt_etal02}. Q0957+561 has been studied by several groups and at different times who each found a single rapid event with durations ranging from less than 1 day to tens of days \citep{schmidt_wambsganss98,gilmerino_etal01,colley_schild03}.  SBS 1520+530 and HE 1104-1805 have both been observed to have multiple events with durations of $\sim$50 days \citep{burud_etal02} and $\sim$1 month \citep{ofek_maoz03}, respectively.  It could be that none of these events are due to lensing by subsolar mass dark matter halos, and larger samples of monitored lens systems will be necessary in order to detect the effects of subsolar mass dark matter halos with any certainty.  

Future surveys may detect large number of quasars lensed by galaxies with sufficient monitoring to measure precise time delays.  The Panoramic Survey Telescope \& Rapid Response System (Pan-STARRS)\footnote{http://pan-starrs.ifa.hawaii.edu.
} and the Large Synoptic Survey Telescope (LSST)\footnote{http://www.lsst.org} are expected to detect $\sim$2000 and $\sim$8000 lenses, respectively \citep{oguri_marshall10}.  Both Pan-STARRS and LSST will perform monitoring and may measure time delays with an accuracy of a few days.  A proposed dedicated lens monitoring mission, Observatory for Multi-Epoch Gravitational Lens Astrophysics \citep[OMEGA;][]{moustakas_etal08} would measure time delays  with an accuracy of $\sim$0.1 days for 100 lenses.  It is important to note that in conjunction with these datasets, new methods of analysis need to be developed, implementing subsolar mass dark matter halos into Bayesian methods of constraining the characteristics of lens systems.

In summary, we present a methodology that can be used to extract the gravitational nanolensing signal that may originate from the presence of subsolar mass dark matter halos in lensing galaxies. We show that such signal may potentially be observable, and that the nanolensing effects can be distinguishable from the microlensing effects caused by stars.
Future surveys will observe a sufficient number of quasar lenses that could be used to undertake a systematic search for the nanolensing signal due to subsolar mass halos. A potential detection of these effects will be invaluable as it will shed light in the nature of dark matter, it will provide information about very early structure formation and will expand our understanding of the energetic sources that power active galactic nuclei.

\begin{acknowledgements}
We thank Alex Geringer-Sameth for useful discussions, Adam Myers and Sean Andrews for help with SDSS quasars, and Kasper Schmidt for sharing his SDSS SQL queries and vital assistance with in creating SDSS quasar samples.  We especially thank Leonidas Moustakas for useful comments and encouragement during the early stages of this work. 

Funding for the SDSS and SDSS-II has been provided by the Alfred P. Sloan Foundation, the Participating Institutions, the National Science Foundation, the U.S. Department of Energy, the National Aeronautics and Space Administration, the Japanese Monbukagakusho, the Max Planck Society, and the Higher Education Funding Council for England. The SDSS Web Site is http://www.sdss.org/.
The SDSS is managed by the Astrophysical Research Consortium for the Participating Institutions. The Participating Institutions are the American Museum of Natural History, Astrophysical Institute Potsdam, University of Basel, University of Cambridge, Case Western Reserve University, University of Chicago, Drexel University, Fermilab, the Institute for Advanced Study, the Japan Participation Group, Johns Hopkins University, the Joint Institute for Nuclear Astrophysics, the Kavli Institute for Particle Astrophysics and Cosmology, the Korean Scientist Group, the Chinese Academy of Sciences (LAMOST), Los Alamos National Laboratory, the Max-Planck-Institute for Astronomy (MPIA), the Max-Planck-Institute for Astrophysics (MPA), New Mexico State University, Ohio State University, University of Pittsburgh, University of Portsmouth, Princeton University, the United States Naval Observatory, and the University of Washington.
\end{acknowledgements}

\bibliography{manuscript}

\end{document}